\documentclass[a4paper,8pt,english]{article}
\usepackage[italian]{babel}
\usepackage[latin1]{inputenc}
\usepackage{latexsym}
\usepackage{natbib}
\usepackage{bbm}
\usepackage{amsmath,amssymb,amsfonts}
\usepackage{graphicx}
\usepackage{graphics}
\usepackage{wrapfig}
\usepackage{amsmath}
\usepackage[a4paper]{geometry}
\begin{document}

\title{\Large  \center \textbf{\textrm{ A stochastic switching control model arising in general OTC contracts with contingent CSA in presence of CVA, collateral and funding}}}
\maketitle
\begin{center}
  October $30$, 2014
\end{center}


\begin{center}
  GIOVANNI MOTTOLA\footnote{Mail: g.mottola@be-tse.it}\\
  Sapienza University of Rome
\end{center}

\begin{center}
\textrm{Abstract}\\
\end{center}
The present work  studies and analyzes  general \emph{defaultable OTC contract} in presence of a \emph{contingent CSA}, which is a theoretical \emph{counterparty risk mitigation mechanism of switching type } that allows the counterparty of a general OTC contract to switch from zero to full/perfect collateralization and switch back whenever she wants until contract maturity paying some switching costs and taking into account the running costs that emerge over time. The  motivation and the underlying economic idea is to show that the current full/partial collateralization mechanisms defined within contracts' CSA - and now imposed by the banking supervision authorities - are ''suboptimal'' and less economic than the \emph{contingent} one that allows to optimally take in account all the relevant driver namely the expected costs of counterparty default losses - represented by the (bilateral) CVA - and the expected collateral and funding costs. In this perspective, we tackle the problem from the \emph{risk management and optimal design } point of view solving - under some working assumptions -  the derived \emph{stochastic switching control model} via \emph{Snell envelope} technique and important results of the theory of the \emph{backward stochastic differential equations with reflection} (RBSDE).
We have also studied the numerical solution providing an algorithm procedure for the value function computation based on an \emph{iterative optimal stopping} approach.

\section{\large Introduction. }

\subsection{Literature review.}

The present research work explores and develops some issues related to the valuation and risk management of general OTC contracts in presence of counterparty risk, namely credit value adjustment (CVA), collateralization and funding.\\
The  counterparty risk valuation and impact in pricing and risk management has gained great importance and relevance in everyday financial markets. Particularly, in the aftermath of the last crisis and Lehman Brothers default, this risk has been regularly traded by the financial institutions on their over the counter positions via the charge of the well-known ''Credit Value Adjustment'' (CVA, we refer to section  two for its formal definition), which was one of the major causes of the relevant losses suffered during the crisis due to  deterioration of the counterparty credit/rating conditions.
Almost at the same time, OTC deals, especially credit derivatives and interest rate swaps, have seen a large increment of the collateral provision and collateralization mechanisms defined within the contract. These are typically defined in a specific annex known as ''\emph{credit support annex}'' (CSA).
The reason for this is that setting a collateralization mechanism allows the counterparties to reduce or even null the CVA (this depends on the type of collateralization, full or partial and on margining), that is the integral of the expected weighted exposures on counterparty default risk taken over the life of the underlying contract or equivalently the cost for the setting of a dynamic hedging strategy to minimize this risk.\\
To remark the relevance of this issue, also the Basel Committee has intervened on  counterparty risk defining new provisions and rules (in \emph{Basel three}) in order to assess and control the counterparty risk. In particular, capital charges have been introduced to take in account for the volatility of the CVA  in addition to a "\emph{central counterparty clearing house}" that operates as ''guarantor'' of all the market transactions with the obligation to set only full collateralysed procedure with the counterparties that transfer to it their credit/default risk.
So counterparty risk  represents a crucial element to assess and analyze.\\
As regards the literature, main recent works have dealt with the modeling of CVA dynamics and collateral mechanisms within specific derivatives products and at portfolio level. As regards this point, are worth of mention the monographic works of   Gregory (2010) and Cesari (2009) which provide a detailed analysis of the problem involved both from the pricing/hedging and risk management point of view. In particular, they tackle the problem related to CVA modeling and hedging highlighting  the possible approaches based on dynamic and static hedging strategies in presence of collateral mechanisms too, focussing especially on operational and computational aspects (at portfolio level in particular) and showing examples and applications to different type of derivative products. Also Bielecki \emph{et al.} (2011) have  faced the CVA hedging issue in a \emph{Markov copula framework} proposing a dynamic hedging strategy based on an abstract contract, called  ''\emph{rolling CDS}'', written on the default event of the counterparty derived via the \emph{mean-variance hedging} approach - given the market incompleteness and the impossibility to replicate the CVA.\\
For what concerns the importance and the role of collateral in credit risk mitigation and CVA reduction, other than the already mentioned works of Cesari (2010), Gregory (2009), are worth of mention the works of  Brigo, Capponi \emph{et al.} (2011) that model and analyze the impact of collateralization (including rehypotecation and netting)  on CVA and on defaultable claims arbitrage-free pricing; Cossin \& Aparicio (2001), working under a \emph{structural framework}, for first propose to use the collateral as a suitable control instrument of the counterparty risk and using it to set a stochastic impulse control problem (with controlled diffusion dynamic) solved  via \emph{variational inequalities} numerical methods.\\
Johannes \& Sundaresan  (2003) instead, analyze the role of collateral in the determination of the market swap rates, highlighting its relevance and impact. Their observations are extended by  Fujii \& Takahashi (2010, 2011) considering the  collateral role in term structure modeling and - adopting  a Duffie \& Huang framework  and using the \emph{ Gauteaux functional derivative} - they are able to show the relevant impact that collateral mechanism defined in both unilateral and bilateral CSA can have on CVA (underlining the relative issues of the cost of funding,  the collateral currency choice problem and the model risk  especially in presence of exotic products).\\
Other works worth of mention on collateral issue are those of Pieterbarg (2010) who introduces it in the classic  Black and Scholes framework and Bielecki, Cialenko \& Iyigunler (2011) which  derive - in a \emph{Markov copula} framework - collateralized CVA dynamic and its impact on arbitrage-free pricing calculating the \emph{spread value adjustment} in the case of CDS products (with all the counterparties involved defaultable).  In relation to funding  imprtant references are those of \emph{Brigo et al.} (2011) that generalizes the CVA valuation in presence of CSA including  funding (defining the so called \emph{FVA}) \emph{Crepey} (2011) that for first rigourously tackles  - using the theory of backward SDE - the \emph{price-hedge problem} for a general contract in presence of CSA and funding.\\

\subsection{Objective and main contributions.}
In the present work we analyze a theoretically  new scheme in which two defaultable counterparties want to define a CSA  characterized by the flexibility and the possibility for both of them to call/put the collateralization  during the life of the underlying claim/contract. We refer specifically to a contingent risk mitigation mechanism that allows the counterparty to switch from zero to full/perfect collateralization (or even partial) and switch back whenever she wants until maturity $T$ paying some \emph{switching costs }and taking into account the\emph{ running costs }that emerge over time.\\
The running costs that we model and consider in the analysis of this problem are - by one side - those related to CVA namely\emph{ counterparty risk hedging costs} and - by the other side - the \emph{collateral and funding/liquidity costs} that emerge when collateralization is active.\\
We can summarize the characteristics and the basic idea underlying the problem  (that we show to admit a natural formulation as a \emph{stochastic (multiple) impulse/switching control problem})   through the so defined \emph{contingent CSA scheme} shown below (Fig. 1.1), in which - by considering also the funding issue in the picture - is present a third party, an \emph{external funder} assumed for convenience \emph{default free} ($\lambda =0$).
\begin{figure}[h!]
  \centering
  \includegraphics[  scale=0.6]{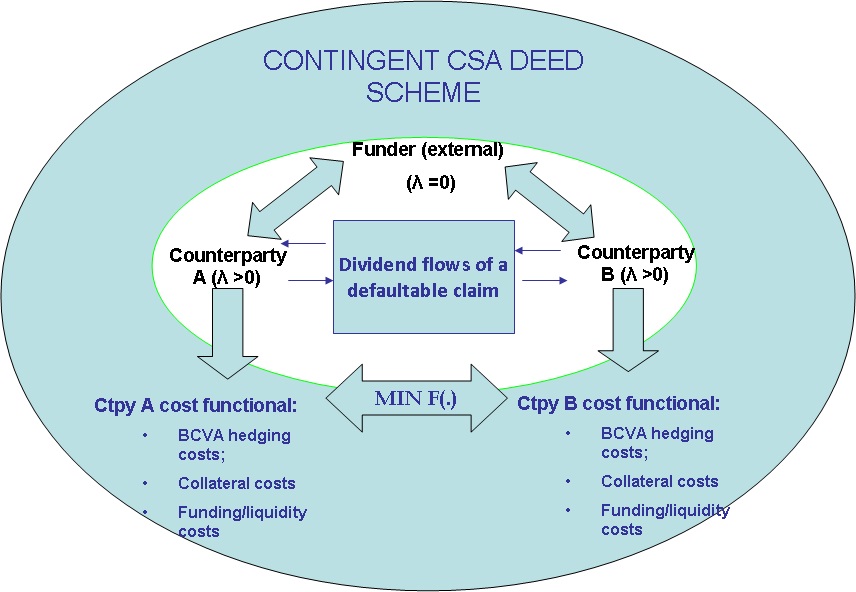}
   \caption{The basic underlying idea. }\label{}
\end{figure}

Our main contributions are a fairly general formulation (not as general as possible) of the problem under analysis as a \emph{stochastic switching control} one (set in \emph{section two an three}) , its solution existence and uniqueness
showed  through suitable \emph{stochastic methods} - mainly the \emph{Snell envelope} and the \emph{reflected backward SDE} representation - (\emph{section four})and the definition of the algorithm for the  numerical solution   based on an \emph{iterative optimal stopping procedure} combined with the \emph{Longstaff-Schwartz method} which we implement in the case of a defaultable interest rate swap (with the contingent CSA that we define) (in \emph{section five}).\\

\section{\large Model framework, Definitions and working Assumptions.}

\subsection{\textit{Model framework.}}
To begin is convenient to describe the framework in which we work and to give some useful definitions  of the processes and variables involved. The framework is the typical one of the \emph{reduced-form models } literature: we are in continuous time and we have a probability space described by the triple $(\Omega, \mathcal{G}_{t}, \mathbb{Q})$ in which lie two strictly positive random time $\tau_{i}$ for $i\in \{A,B\}$, which represent the \emph{default times} of the counterparties considered in our model.  In addition, we define the \emph{default process } $H^{i}_{t} = \mathbbm{1}_{\{\tau_{i} \leq t\}}$ and the relative filtration $\mathbb{H}^{i}$ generated by $H_{t}^{i}$ for any $t\in \mathbb{R}^{+}$. This implies that the \emph{full filtration} of the model is given by $\mathbb{G} = \mathbb{F} \vee \mathbb{H}^{A} \vee \mathbb{H}^{B}$ where $\mathbb{F}$ is the (risk-free) \emph{market filtration} usually generated by a Brownian motion $W$ (or a vector $\hat{W}$) adapted to it, under the real measure $\mathbb{Q}$. In the rest of the work we are used to denote with $\mathcal{G}_{t} = \sigma(\mathcal{F}_{t} \vee \mathcal{H}^{A}\vee\mathcal{H}^{B})_{t\geq0}$ that is the sigma algebra of the enlarged market filtration. In addition, we remember that all the processes we consider, in particular $H^{i}$,  are \emph{c\'adl\'ag semimartingales} $\mathbb{G}$ adapted and $\tau^{i}$ are $\mathbb{G}$ stopping times (which means to be in a \emph{Skorohod topology}).\\
For convenience, let us define the first default time of the counterparties as $\tau = \tau_{A}\wedge\tau_{B}  $ which also represent the ending/exstinction time of the underlying contract, with the corresponding indicator process $H_{t} = \mathbbm{1}_{\{\tau\leq t\}}$.
For what concerns the underlying market model it is assumed arbitrage-free, meaning that it admits a \emph{spot martingale measure} $\mathbb{Q}^{*}$ (not necessarily unique) equivalent to $\mathbb{Q}$. A spot martingale measure is associated
with the choice of the savings account $B_{t}$ (so that  $B^{-1}$ as discount factor) as a numeraire that, as usual,  is given by $\mathcal{F}_{t}$-predictable process
\begin{equation}
    B_{t}= \exp \int_{0}^{t}r_{s}ds, \:\:\forall \: t \in \mathbb{R}^{+}
\end{equation}
For convenience, we assume that the \emph{immersion property} holds in our framework, so that every  c\'adl\'ag $\mathbb{G}$-adapted (square-integrable) process is also $\mathbb{F}$-adapted. In particular, the processes and random variable that we encounter in the following sections will be in general $\mathcal{G}$-adapted or predictable and often $\mathcal{G}_{\tau}$-measurable, but because the default time $\tau$ is inaccessible in the reduced framework, one needs to work with $\mathcal{F}$-adapted or predictable process  called \emph{pre-default value } processes, denoted with $X_{-}$ that is the left limiting process of $X$. The following lemma is a classic results\footnote{See Jeanblanc \& Le cam (2008) for details.} that allows this change of filtration. \\

\textbf{Lemma 2.1.1.} \emph{Set $J := 1-H = \mathbbm{1}_{\{t\leq \tau \}}$.For any $\mathcal{G}$-adapted, respectively $\mathcal{G}$-predictable process $X$ over $[0,T]$, there exists a unique $\mathcal{F}$-adapted,  respectively $\mathcal{F}$-predictable,  process $\tilde{X}$ over $[0,T]$, called the pre-default value process of $X$, such that $JX = J\tilde{X}$, respectively $J_{-}X = J_{-}\tilde{X}$.}\\

\textbf{Remarks 2.1.1.} 1) In particular, in our framework we assume also true that, $\mathcal{F}$-local martingales stopped at $\tau$ are also $\mathcal{G}$-local martingale, which is a just a modification of the immersion hypothesis.\\
2) Therefore is worth to remind that $\mathcal{F}$-adapted c\'adl\'ag process cannot jump at $\tau$ so one has that $\Delta X = 0$ almost
surely, for every $\mathcal{F}$-adapted c\'adl\'ag  process $X$.\\
These are technical conditions necessary to make more workable the framework and the model that being characterized by defaultable  processes, whose value at default is inaccessible,  allow to  set off jump behavior and all the problems related to consider in the model also \emph{close out clause}  and the contract \emph{replacement costs} (for which we refer  for example to Brigo, Capponi et al (2011)). Anyway, this will be clear in the next section.\\

\subsection{ \textit{Main definitions.}}

Let us firstly recall the main definitions that we need for our model about bilateral CVA, CSA with perfect/full collateralisation and our theoretical contingent CSA scheme.
Further details can be found in (reference).\\
We start giving some general definitions of the price process (or NPV) and CVA for a defaultable claim for which no CSA that is no collateralization and other mitigation mechanism has been defined by the parties within the relative contract. \\
First of all, by a defaultable claim signed between two risky counterparty maturing at time $T$ we mean the quadruple $(X; \mathbf{A};Z; \tau )$, where $X$ is an $\mathcal{F}_{T}$-measurable random variable, $\mathbf{A} = (\mathbf{A}_{t})_{t\in[0,T ]}$ is an $\mathbb{F}$-adapted, continuous process of finite variation with $\mathbf{A}_{0}= 0$, $Z = (Z_{t})_{t\in[0,T ]} $ is an $\mathbb{F}$-predictable process, and $\tau=\tau_{A}\wedge \tau_{B}$ is the first default time of one of the counterparties (assumed completely inaccessible in the reduced form  framework).\\

\textbf{Definition 2.2.1 (Clean dividend and price process).}
\emph{The clean dividend process $D^{rf}_{t}$ of a counterparty default free (exchange traded)  contract is the $\mathcal{F}$-adapted process described by the final payoff $X$, the cashflows $A$ and $\tau =\tau^{i}=\infty$ as follows
\begin{eqnarray}
  D_{t}^{rf} &=& X  \mathbbm{1}_{[T, \infty]}(t) + \sum_{i\in\{A,B\}}\bigg(\int_{]t, T]} d\mathbf{A}^{i}_{u} \bigg)  \;\;t\in[0,T].
\end{eqnarray}
The clean price process $S^{rf}$ would be simply represented by the integral over time of the dividend process under the relative pricing measure, that is
\begin{equation}
   S_{t}^{rf}= B_{t}\mathbb{E}_{\mathbb{Q}^{*}} \bigg( \int_{]t,T]} B_{u}^{-1} dD^{rf}_{u} \big| \mathcal{F}_{t} \bigg)\;\;t\in[0,T].
\end{equation}}


\textbf{Definition 2.2.2 (Bilateral risky dividend and price process).} \emph{The dividend process $D_{t}$ of a defaultable claim with bilateral counterparty risk $(X; \mathbf{A};Z; \tau )$, is defined as
the total cash flows of the claim until maturity $T$
that is formally
\begin{eqnarray}
  D_{t} &=& X \mathbbm{1}_{\{T<\tau\}} \mathbbm{1}_{[T, \infty]}(t) + \sum_{i\in\{A,B\}}\bigg(\int_{]t, T]}  (1-H_{u}^{i})d\mathbf{A}^{i}_{u} +  \int_{]t,T] } Z_{u}dH_{u}^{i}\bigg)  \;\;t\in[0,T]
\end{eqnarray}
for $i \in \{A,B\}$.\\
Similarly, the (ex-dividend) price process $S_{t}$ of a defaultable claim with bilateral counterparty risk maturing in $T$ is defined as the integral of the discounted  dividend flow under  the risk neutral measure $\mathbb{Q}^{*}$, namely the $NPV_{t}$ of the claim, that is formally
\begin{equation}
   NPV_{t} = S_{t}= B_{t}\mathbb{E}_{\mathbb{Q}^{*}} \bigg( \int_{]t,T]} B_{u}^{-1} dD_{u} \big| \mathcal{G}_{t} \bigg)\;\;t\in[0,T].
\end{equation}}

Given these definitions for the price and dividend process, we are able to state the following general definition for the \emph{bilateral CVA} (at a given time $t$)
\begin{equation*}
    BCVA_{t} = S_{t}^{rf} - S_{t} \;\: \forall\: t \in  [0,T].
\end{equation*}
as the difference between the counterparty risk free and risky price process/NPV of the contract.
Let us state the following proposition on bilteral CVA.\\

\textbf{Proposition 2.2.3 (Bilateral CVA).} \emph{The bilateral CVA process of a defaultable claim with bilateral counterparty risk $(X; \mathbf{A};Z; \tau )$ maturing in $T$ satisfies the following relation\footnote{The formulation is seen from the point of view of $B$. Being symmetrical between the party, just the signs change.}
\begin{eqnarray}
  BCVA_{t} &=& CVA_{t} - DVA_{t} \nonumber \\
  &=& B_{t}\mathbb{E}_{\mathbb{Q}^{*}} \bigg[ \mathbbm{1}_{\{t< \tau=\tau_{B}\leq T\}}B_{\tau}^{-1}(1-R_{c}^{B})(S^{rf}_{\tau})^{-} \bigg|\mathcal{G}_{t} \bigg] + \nonumber \\
  &-& B_{t}\mathbb{E}_{\mathbb{Q}^{*}}\bigg[ \mathbbm{1}_{\{t< \tau=\tau_{A}\leq T\}}B_{\tau}^{-1}(1-R_{c}^{A})(S^{rf}_{\tau})^{+}\bigg|\mathcal{G}_{t}  \bigg]
\end{eqnarray}
for every $t \in[0,T]$, where $R_{c}^{i}$ for $i\in\{A,B\}$ is the counterparty recovery rate (process).}\\

\emph{Proof.} The proposition proof follows the same simplified lines
of the proposition 2.9 of Bielecki, Cialenko \& Iyigunler (2011).\\

Let us remark (see for example Brigo \emph{et al. }(2010)) that in case of independence between discount factor and default intensities\footnote{This orthogonality hypothesis will be useful in the implementation and numerical part.} the CVA
process can be represented in terms of EPE and ENE as follows
\begin{eqnarray}
  BCVA_{t} &=&B_{t}\mathbb{E}_{\mathbb{Q}^{*}}\big( \mathbbm{1}_{\{t< \tau=\tau_{A}\leq T\}}B_{\tau}^{-1}(1-R_{c}^{A})(S^{rf}_{\tau})^{+}  \big)
  - B_{t}\mathbb{E}_{\mathbb{Q}^{*}} \big( \mathbbm{1}_{\{t< \tau=\tau_{B}\leq T\}}B_{\tau}^{-1}(1-R_{c}^{B})(S^{rf}_{\tau})^{-} \big) \nonumber \\
  &=& B_{t} \int_{]t,T]}B_{s}^{-1}EPE_{s}\mathbb{Q}^{*}(\tau_{A}\in ds,s\leq \tau_{B})-B_{t} \int_{]t,T]}B_{s}^{-1}ENE_{s}\mathbb{Q}^{*}(\tau_{B}\in ds,s\leq \tau_{A})
\end{eqnarray}
for $\; t\in[0,T]$ where from the framework section we know that  $\mathbb{Q}^{*}(\tau_{i}\in ds,s\leq \tau_{i}) =\lambda^{i}_{t}:= G^{-1}(t) \mathbb{Q}^{*} (t=\tau_{i}  \in dt)$, namely the counterparty default intensities.\\


As regards the CSA, it can be described, in general terms, as an articulated part of the OTC contracts devoted to mitigate the counterparty risk through the definition of a collateral agreement between the parties over the life of the underlying claim\footnote{Now, thanks to diffusion of central clearings to mitigate the counterparty risk, this agreement has been considering, as a third party involved, the central clearing as a guarantee of the transaction.}. It has been largely used also to reduce the capital requirements imposed by the Basel committee on portfolio exposures and to give a more competitive price to the CVA charges.\\
As already mentioned, this type of agreement usually contains a lot of factors, clauses and events that are complex to model (see Gregory (2011)  for details).
Let us now generalize the definition of CVA in presence of the CSA.\\  
\textbf{Definition 2.2.4 (Collateral account/process)}. \emph{Let us define for the bilateral CSA of a contract the positive/negative threshold  with $\Gamma_{i}$ for $ i=\{A,B\}$ and the  positive minimum transfer amount with $MTA$. The collateral process $Coll_{t} : [0,T]\rightarrow \mathbb{R}$ is a stochastic  $\mathcal{F}_{t}$-adapted process\footnote{In the literature, depending on the CSA provisions and modeling choice, the process is also considered $\mathcal{F}_{t}$-predictable or in general adapted to $\mathcal{G}_{t}$.} defined as follows
\begin{equation}
    Coll_{t}=\mathbbm{1}_{\{S^{rf}_{t} > \Gamma_{B} + MTA\}}(S^{rf}_{t} - \Gamma_{B}) + \mathbbm{1}_{\{S^{rf}_{t} < \Gamma_{A} - MTA\}}(S^{rf}_{t} - \Gamma_{A}),
\end{equation}
on the time set $ \{t<\tau\}$, and
    \begin{equation}
    Coll_{t}=\mathbbm{1}_{\{S^{rf}_{\tau^{-}} > \Gamma_{B} + MTA\}}(S^{rf}_{\tau^{-}} - \Gamma_{B}) + \mathbbm{1}_{\{S^{rf}_{\tau^{-}} < \Gamma_{A} - MTA\}}(S^{rf}_{\tau^{-}} - \Gamma_{A}), \;
\end{equation}}
 on the set $ \{\tau\leq t <\tau +\Delta t\}$.\\

\textbf{Definition 2.2.5 (CSA close-out cashflows)}.\emph{ Let us define the recovery rate $R_{c}^{i}$ for $i \in\{ A,B\}$ a $\mathcal{G}$-predictable process set as a positive constant and the uncollateralized mark to market as $\theta_{t}= S^{rf}_{t} - Coll_{t}$ (for $t=\tau)$. The CSA close-out cashflows is the real valued left limited $\mathcal{G}_{\tau^{-}}$-measurable process defined as follows
\begin{equation*}
    Cf^{CSA}_{t}= Coll_{t}+   \mathbbm{1}_{\{t= \tau_{B}\}} (  R_{c}^{B}\theta_{t}^{+} -\theta_{t}^{-})- \mathbbm{1}_{\{t= \tau_{A}\}} (R_{c}^{A}\theta_{t}^{-} - \theta_{t}^{+} ) - \mathbbm{1}_{\{t= \tau_{A}= \tau_{B}\}} \theta_{t}  ).
\end{equation*}}


By using these definitions and following  the same line  of proof the proposition 2.2.3. this proposition states.\\

\textbf{Proposition 2.2.6 (CVA (bilateral) with CSA)}.\emph{
The bilateral CVA for a defaultable claim  maturing in $T$ and mitigated by a partial collateralization satisfies the following relation
\begin{eqnarray}
   BCVA_{t}^{CSA} &=& B_{t} \mathbb{E}^{\mathbb{Q}^{*}}\bigg[  \mathbbm{1}_{\{t<\tau= \tau_{B}\leq T\}}B_{\tau}^{-1}(1-R_{c}^{B})(S_{\tau}^{rf} - Coll_{\tau})^{-} \big| \mathcal{G}_{t}\bigg] \nonumber  \\
   &-& B_{t} \mathbb{E}^{\mathbb{Q}^{*}}\bigg[ \mathbbm{1}_{\{t<\tau= \tau_{A}\leq T\}}B_{\tau}^{-1}(1-R_{c}^{A})(S_{\tau}^{rf} - Coll_{\tau})^{+} \big| \mathcal{G}_{t} \bigg] \;
\end{eqnarray}
$\forall t\in[0,T]$, where the first term is the (unilateral) CVA and the second the DVA (collateralised).}\\


In the case  of \textbf{\emph{CSA with perfect/full collateralization}}, say $Coll^{Perf}_{t}$, collateralization can be easily understood as the limit case of the partial collateralization  when the delta between the margin dates, say $\Delta_{t_{m}} = t_{m}- t_{m-1}$, tends to zero  and no thresholds and minimum transfer amount are defined in the related CSA. This means that $Coll^{Perf}_{t}$ is always equal to the mark to market, namely to the (default free) price process $S_{t}^{rf}$ of the underlying claim, that is formally (by taking the definition 2.2.4)
\begin{equation}
    Coll_{t}^{Perf}=\mathbbm{1}_{\{S^{rf}_{t} > 0 \}}(S^{rf}_{t} - 0) + \mathbbm{1}_{\{S_{t}^{rf} <0\}}(S^{rf}_{t} - 0) = S^{rf}_{t} \;\; \forall \:  t \in [0,T], \; on\:\{t<\tau\}.
\end{equation}
and
\begin{equation}
    Coll_{t}^{Perf}=  S^{rf}_{\tau^{-}} \;\; \forall \:  t \in [0,T], on \: \{ \tau\leq t <\tau+\delta t\}
\end{equation}

Then, by plugging the last relation in the  definition 2.2.5 of CSA cashflows, one easily gets

\begin{equation}
    Cf^{CSA}_{t}= Coll^{Perf}_{t} \;\; \forall \:  t = \tau \in [0,T].
\end{equation}

Actually, the CSA close out cashflows won't get sense anymore because the counterparty risk is zeroed with perfect collateral and, in fact, the CVA defined in 2.2.3 become
\begin{equation}
     BCVA_{t}^{Coll^{Perf}} = B_{t}E_{\mathbb{Q}^{*}}\bigg[ \mathbbm{1}_{\{t<\tau_{B}\leq T \}} (0) | \mathcal{G}_{t} \bigg] - B_{t}E_{\mathbb{Q}^{*}}\bigg[ \mathbbm{1}_{\{t<\tau_{A}\leq T \}} (0) | \mathcal{G}_{t} \bigg] =0  \;\; \forall t\in[0,T]
\end{equation}
that implies
\begin{eqnarray}
  BCVA_{t}^{Coll^{Perf}} &=& S_{t}^{rf} - S_{t} =0\Longrightarrow \nonumber \\
   S_{t}&=&  S_{t}^{rf}  \; \forall t\in [0,T]
\end{eqnarray}

namely the equality between the defaultable price and risk free price of the claim.\\


Now we are ready to define the contingent mechanism of switching type that we assume to be set between the parties in our model. It can be seen as a mix of zero and perfect collateralization and it is natural to model it  (as will be clearer from the next section) introducing the switching indicators $z_{j}$ and times $\tau_{j}$ for $j=1,\dots,M$, that affects - being the agent controls - the system dynamic and consequently the processes involved, in particular the collateral process. Let us consider just two possible switching regimes (but it can be easily generalized also to the case of partial collateralization) so that the indicator will take only two values, say $z_{j}=\{0,1\}$:\\
\begin{itemize}
  \item  when $z_{j}=1$ the collateral is null, no collateralization is active, that implies a full BCVA (with no CSA);

  \item while for $z_{j}=0$ we have a CSA with perfect/full collateralization (which zeroes the BCVA).

\end{itemize}
So,  as done before we give the main definitions for collateral and CVA process which now depend also on these switching control $\{\tau_{j}, z_{j}\}_{j=1}^{m}$ variables that needs to be determined optimally. How, it will be the object of the following section.\\

\textbf{Definition 2.2.7 (Contingent Collateral process)}.\emph{The contingent collateral  $Coll^{C}_{t}$ is the $\mathcal{F}_{t}$-adapted process defined for any time $t\in[0,T]$ and  for every switching time $\tau_{j} \in [0,T] $ and $j=1,\dots,M$, switching indicator $z_{j}$ and default time $\tau$ (defined above as $\min\{\tau_{A},\tau_{B}\} $ ), as follows
\begin{equation}
    Coll^{C}_{t} = S^{rf}_{t}\mathbbm{1}_{\{ z_{j}=0 \}} \mathbbm{1}_{\{ \tau_{j} \leq t < \tau_{j+1} \}} + 0 \mathbbm{1}_{\{ z_{j}=1 \}} \mathbbm{1}_{\{ \tau_{j} \leq t < \tau_{j+1} \}} \;\: on \: \{ t<\tau\}
\end{equation}
on the set $ \{ t<\tau\}$, and
\begin{equation}
    Coll^{C}_{t} = S^{rf}_{\tau^{-}}\mathbbm{1}_{\{ z_{j}=0 \}} \mathbbm{1}_{\{ \tau_{j} \leq t < \tau_{j+1} \}} + 0 \mathbbm{1}_{\{ z_{j}=1 \}} \mathbbm{1}_{\{ \tau_{j} \leq t < \tau_{j+1} \}} \;
\end{equation}
on the set  $\{ \tau\leq t <\tau+\Delta t\}$ where we recall that $S_{t} = Coll_{t}^{Perf} $ from point $b)$ results.}\\

Using Definition 2.2.7 and the results stated for close out amounts and bilateral CVA, we get easily the following results. Firstly, setting $D^{C}$ and $S^{C}$ as the  dividend and price process in presence of the contingent CSA just defined, we can set
\begin{eqnarray*}
  D^{C}_{t} &=&  D^{rf}_{t}\mathbbm{1}_{\{ z_{j}=0 \}} \mathbbm{1}_{\{ \tau_{j} \leq t < \tau_{j+1} \}} + D_{t} \mathbbm{1}_{\{ z_{j}=1 \}} \mathbbm{1}_{\{ \tau_{j} \leq t < \tau_{j+1} \}} \\
  S^{C}_{t} &=&  S^{rf}_{t}\mathbbm{1}_{\{ z_{j}=0 \}} \mathbbm{1}_{\{ \tau_{j} \leq t < \tau_{j+1} \}} + S_{t} \mathbbm{1}_{\{ z_{j}=1 \}} \mathbbm{1}_{\{ \tau_{j} \leq t < \tau_{j+1} \}}
\end{eqnarray*}

for $t \in[0,T\wedge\tau]$. From these relations  it is easy to recover the general formulation for the bilateral CVA process in the contingent case as follows
\begin{eqnarray*}
  BCVA^{C}_{t} &=& S^{rf}_{t} - S^{C}_{t} \\
   &=& S^{rf}_{t} - (S^{rf}_{t}\mathbbm{1}_{\{ z_{j}=0 \}} \mathbbm{1}_{\{ \tau_{j} \leq t < \tau_{j+1} \}} + S_{t} \mathbbm{1}_{\{ z_{j}=1 \}} \mathbbm{1}_{\{ \tau_{j} \leq t < \tau_{j+1} \}}) \\
   &=&  0 \mathbbm{1}_{\{ z_{j}=0 \}} \mathbbm{1}_{\{ \tau_{j} \leq t < \tau_{j+1} \}} + BCVA_{t}\mathbbm{1}_{\{ z_{j}=1 \}} \mathbbm{1}_{\{ \tau_{j} \leq t < \tau_{j+1} \}}
\end{eqnarray*}
so that the following definition is well posed.\\

\textbf{Definition 2.2.7 (BCVA with contingent CSA)}. \emph{The bilateral CVA of a contract with contingent CSA of switching type, $BCVA^{C}_{t}$ is the $\mathcal{G}_{t}$-adapted process defined for any time $t\in[0,T]$,  for every switching time $\tau_{j} \in [0,T] $ and $j=1,\dots,M$, switching indicator $z_{j}$ and default time $\tau$ (defined as above), as follows
\begin{equation}
    BCVA^{C}_{t} = BCVA_{t}\mathbbm{1}_{\{ z_{j}=1 \}} \mathbbm{1}_{\{ \tau_{j} \leq t < \tau_{j+1} \}} + 0 \mathbbm{1}_{\{ z_{j}=0 \}} \mathbbm{1}_{\{ \tau_{j} \leq t < \tau_{j+1} \}}, \;\: for \: t\in[ 0,T\wedge\tau],
\end{equation}
where the expression for $BCVA$ is known from Proposition 2.2.3.}\\

We end by noting that the CSA closeout cashflows are different from zero only when collateralization is active, that is
\begin{equation*}
    Cf^{CSA}_{t}= Coll^{C}_{\tau} \;\; \forall \:  t = \tau \in [0,T].
\end{equation*}

\textbf{Remarks 2.2.8.} We underline that all these processes are in general c\'adl\'ag semi-martingales and is difficult  to deal with their dynamics which are also recursive and non linear being affected by agent switching controls. In the next section we will see from the set up of the stochastic control the working assumptions needed to ease the problem analysis.\\

\subsection{ \textit{Funding with CVA and collateral.}}

The funding issue and how to model it in the whole issue of pricing and hedging of derivatives and defaultable claims in particular,  is one of the main topic in the literature (a clear analysis of the issue can be found in Brigo \emph{et al.} (2011)). 
As concerns our problem, we assume the existence of a \emph{cash account} hold by an external funder, say $C^{Fund}_{t}\gtrless 0$, that can be positive or negative depending on the funding ($>0$) or investing ($<0$) strategy that counterparty uses in relation to the price and hedge of the specific deal which is also collateralysed in our case.  Including the funding in the analysis, this has impact on:
\begin{itemize}
  \item the cashflows related to the hedging portfolio/strategy of the specific deal, especially if the hedge cannot be perfect as in the case of defaultable claims so that the trader needs to fund or invest the cash surplus deriving from his hedging portfolio;
  \item the cashflows related to CVA hedging; this is usually included in the hedging portfolio of the whole claim;
  \item the cashflows related to the collateral/CSA defined between the parties; for example, when the exposure trespass a specified threshold and the counterparty has to post/receive the collateral it incurs in funding/opportunity costs.
\end{itemize}
In order to reduce the problem of recursion in  relation to CVA, we make the assumption that the CVA value is charged to the counterparty but no hedging portfolio is set, so that the funding costs are null $C^{Fund} = 0$ (namely the funding account is not active)\footnote{An alternative simplifying assumption is that the CVA hedging is realized without resorting to funding. }.\\
As regards the funding issue in  relation to the collateralization, under the assumptions of segregation (no rehypo) and collateral made up by cash, we distinguish the following cases:
\begin{description}
  \item[1)] If the counterparty has to post collateral in the margin account, she sustains a funding cost, applied by the external funder, represented by the \emph{borrowing rate} $r^{borr}_{t} = r_{t} + s_{t}$ that is the risk free rate plus a credit spread (s) (that is usually different from the other party) . By the other side, the counterparty receives by the funder  the remuneration on the collateral post, that is usually defined in the CSA as a risk free rate plus some basis points, so that we can approximate it at the risk free rate, that is $r_{t} + bp_{t} \cong r_{t}$.\\
      Clearly, we remark that the existence of the above defined rate implies the existence of the following funding assets
      \begin{eqnarray}
         dB^{borr}_{t} &=& (r_{t} + s_{t})B^{borr}_{t}dt  \\
        dB^{rem}_{t} &=& (r_{t}+bp_{t} )B^{rem}_{t}dt
      \end{eqnarray}
  \item[2)] Instead, let's consider the counterparty that call the collateral: as above the collateral is remunerated at the rate given (as the remuneration for the two parties can be different) by $\bar{B}^{rem}$, but she cannot use or invest the collateral amount (that is segregated), so she  sustains an \emph{opportunity cost}, that can be represented by the rate $r^{opp}_{t} = r_{t} + \pi_{t}$, where $\pi$ is a premium over the risk free rate. \\
      Hence, we assume the existence of the following assets too
      \begin{eqnarray}
         dB^{opp}_{t} &=& (r_{t} + \pi_{t})B^{opp}_{t}dt  \\
        d\bar{B}^{rem}_{t} &=& (r_{t}+\bar{bp}_{t} )\bar{B}^{rem}_{t}dt
      \end{eqnarray}
\end{description}
The objective of the last definitions will be  clearer from the next subsection in which we pass to define the counterparty objective and in the third section in which we formally define the stochastic switching control problem that we want to tackle.

\subsection{\textit{Counterparty  objective and recap of the working assumptions.}}

Now we have in mind all the relevant elements and variables that the counterparty of a collateralysed defaultable claim of contingent type has to take in consideration to  define the  optimality criterion/objective. As already said, the counterparty is assumed to be counterparty risk averse, she wants to minimize the negative impact of CVA, namely the costs of the default of the other party, via the definition of a contingent collateral/CSA,  but she wants to do it optimally  deciding the optimal times when to switch to a partial or full collateralization. Of course the collateralization implies other running costs, like the funding ones. So, the objective will be to determine over the set of controls represented by the switching time (and indicator) the optimal switching strategy that minimizes over $[0,T]$ the overall running costs related to CVA, collateral and funding  and the instantaneous switching costs that we need to model. This kind of problem is highly non linear and recursive, as already highlighted above, and because of in the partial collateralization case the CVA process is complicated by the presence of the CSA close out cash-flows that depends recursively on the controls (the switching times), this makes the analysis more complex. So, for the next sections, we leave aside the partial collateralization and we analyze the problem for the counterparty in the twofold decision case: \emph{''switch from zero collateral (full CVA) to full collateralization (zero CVA)'' (and viceversa)}.\\
We recall here the main working assumption that we use in the analysis of the problem for the following sections:
\begin{description}
  \item[\textbf{Hp 1)}] \emph{The analysis is done under the ''\emph{symmetry hypothesis''}: the parties of the deal have symmetric objectives and similar default intensities 
      in order to reduce problem dimension (this is a working assumption particularly useful in the numerical part).}
  \item[\textbf{Hp 2)}] \emph{Both the parties are counterparty risk averse but we assume no strategic interaction (issue that will be object of an other work).}
  \item[\textbf{Hp 3)}] \emph{The analysis is focussed on the full/perfect collateralization case;}
  \item[\textbf{Hp 4)}] \emph{The CVA cost process is not funded,   $C^{Fund} = 0$;}
  \item[\textbf{Hp 5)}] \emph{All the processes considered are intended to be pre-default value processes (as from lemma 2.1.1.)}
\end{description}

\section{\large Stochastic Switching Control Problem:  Formulation and Solution Existence.}

\subsection{ \textit{ Model dynamics and controls.}}

From the reasoning and definitions of the last section is quite clear what kind of stochastic control problem we are facing: we are (conterparty $A$) in $t=0$ and we want to determine the optimal strategy that minimizes the expected costs  deriving from switching the collateral from zero to full  and viceversa until the maturity of the underlying defaultable claim. It is natural to formalize this problem  as a \emph{multiple switching control problem with finite horizon} (a particular case of the more general class of \emph{impulse control models}) whose main ingredients  are:
\begin{enumerate}
  \item the setting of the \emph{objective functional} which is generally made up of an objective function, a reward function and a switching/impulse cost function;
  \item the setting of the \emph{stochastic dynamics} that describe the state of the system considered;
  \item the setting and modeling of the \emph{controls set} in which search for the solution of the problem.
\end{enumerate}
From the list set above, the last point is  strictly related to the model for the ''system'' dynamic, which is more problematic given that the natural choice would be to consider the CVA dynamic as it enters the objective functional of counterparty $A$ (and symmetrically $B$ as discussed above). But it can be shown\footnote{See, for example, Bielecki, Cialenko \& Iyigunler (2011).} that  CVA dynamic is quite complicate and is not an \emph{Ito diffusion}, so we need to use different solution methods (for general c\'adl\'ag processes) that we leave for further research. Here, we we use - at cost of a minor generalization -\\
 the bilateral CVA expression set in the latter section under the hypothesis of interest rate orthogonal to  $\tau$, namely the default intensities of counterparty $B$. So in the following, for CVA/BCVA we refer to the formula set in the equation 2.7, that we recall here as follows
\begin{eqnarray}
  BCVA_{t}  &=& B_{t} \int_{]t,T]}B_{s}^{-1} \mathbb{E}\big( (1-R_{c}^{A})(NPV_{\tau})^{-}|\tau_{A} =s,\tau_{A}\leq \tau_{B}  \big)\lambda^{A}_{s}ds \nonumber \\
  &-& B_{t} \int_{]t,T]}B_{s}^{-1} \mathbb{E}\big( (1-R_{c}^{B})(NPV_{\tau})^{+}|\tau_{B} =s,\tau_{B}\leq \tau_{A}  \big)\lambda^{B}_{s}ds
\end{eqnarray}
where we have used the default intensity definition for $\lambda_{t}^{i}$ and we have made an abuse of notation given the price process definition $S_{t }^{rf}:= NPV_{t}$.\\
In order to model the system dynamic in our model, let us note that the BCVA, for a given process realization $(\omega\in \Omega)$ (and varying $t\leq T$ ), can be defined as a \emph{bounded} and $\mathcal{F}_{t}$-\emph{measurable} function $f(.)$:
\begin{equation*}
    BCVA(t)=:f(t, NPV, \lambda, B_{t}, \alpha) = f(t, X, \lambda, \alpha)
\end{equation*}
of
\begin{description}
  \item[-] the price process (default free) or NPV, that can be expressed as a function $h(t,X,k) $ (also $\mathcal{F}_{t}$-\emph{measurable} )  of time, of some constants ($k$) and a stochastic factor $X$\footnote{This can be in general a stochastic vector.}, that is typically the interest rate process, for example in an interest rate swap;
  \item[-] of the default process $\tau$ which is determined through the intensity process, say $\lambda$ that for convenience (in order to reduce problem dimensionality) we have assumed similar for both the counterparties $\lambda_{t}^{A}\approx \lambda_{t}^{B}$;
  \item[-] discount factors that presume the existence in the model of a bank account defined by $B_{t}^{-1}= \exp(-\int_{0}^{t}r_{s}ds)$;
  \item[-] a vector of constant factors $\alpha=(k,R_{c})$ defined in the contract, like the recovery rate $R_{c}=1-LGD_{c}$ or the ''strike rate'' that is included in $k$.
\end{description}
So, we are left to choose a model for   $(X,\lambda)$ namely the interest rates and the default intensities, for which we assume the following general markovian diffusions
\begin{eqnarray}
  dX_{t} &=& \mu(t,X_{t})X(t) dt + \sigma(t,X_{t}) X(t) dW_{t}^{x};\quad X(0)=x_{0}\\
  d\lambda_{t} &=& \gamma(t,\lambda_{t})\lambda(t)dt + \nu(t,\lambda_{t}) \lambda(t)dW_{t}^{\lambda}; \quad \lambda(0)=\lambda_{0} \\
  \quad d\langle X, \lambda\rangle_{t} &=&   d\langle W_{t}^{x}, W_{t}^{\lambda} \rangle_{t}=\rho_{X,\lambda} dt. \nonumber
\end{eqnarray}
We keep this in mind and we remark that the dynamics of the system we are considering is affected by the counterparty choice to switch from zero to full/partial collateral or viceversa, so we need to define the control set for our problem.\\
In a multiple switching/impulse control type problem (with finite horizon) like this, in which  counterparty $A$  can intervene or ''give impulse'' to the system every times in $[0,T]$ through a sequence of switching times, say $\tau_{j} \in \mathcal{T}$, with $\mathcal{T}\subset [0,T]$, where we denote the switching set
\begin{equation*}
    \mathcal{T}:= \big\{ \tau_{1},\dots, \tau_{M} \} = \big\{ \tau_{j} \big\}_{j=1}^{M}
\end{equation*}
with the last switching $\{\tau_{M} \leq T\}$ ($M< \infty$) and the $\tau_{j}$ are, by definition of \emph{stopping times}, $\mathcal{F}_{t}$-measurable random variable. Therefore, at any of these switching times, we need to define the related set of the \emph{switching/impulse indicator}, which is quite trivial in our case with only two possible switching regime:
\begin{equation*}
    \mathcal{Z}:= \big\{ z_{1},\dots, z_{M}\} = \big\{ z_{j}\big\}_{j=1}^{M}
\end{equation*}
with the  $z_{j}$'s that are  $\mathcal{F}_{\tau_{j}}$-measurable switching indicators taking values $z_{j}=\{ 0,1\} \: \forall j=1,\dots,M $, so we have
\[
  \begin{cases}
        z_{j}= 1  \Rightarrow  & ''zero collateral'' (full CVA) \\
         z_{j}= 0  \Rightarrow  & ''full collateral'' (null CVA)
        \end{cases}
\]\\
Here we are focussing on the simpler case of collateralization leaving the ''partial case'' aside.
So we have that the control set for our problem is formed of all the sequences of indicators and switching times $ \mathcal{C} \in \mathbb{R}_{M}^{+}$, that is
\begin{equation*}
    \mathcal{C}=\big\{ \mathcal{T}, \mathcal{Z} \big\} = \big\{ \tau_{j}, z_{j}\big\}_{j=1}^{M}
\end{equation*}
with the convention - and without losing generality - that in $t=0$ the initial condition of the indicator be $z_{0}=1$ because usually contracts start with no collateralization at the inception\footnote{In fact, we could have assumed a contract starting with full collateralization $z_{0}=0$ at inception in $t=0$, but the analysis would have been the same.   }.\\

\textbf{Remarks 3.1.1.} It is worth of mention that, as defined above, the set of switching controls can be made up of maximum $M$ switching times $\tau_{j}$ (and related indicators $z_{j}$) being the counterparty who decides optimally the number of switches that minimize her cost functional. In general one could leave unbounded the number of switched allowed, namely $M=\infty$ leaving its number determined endogenously by the solution of the stochastic optimization. An other more restrictive possibility that is quite common in CSA contracts is to allow switching only at a given set of dates $\tau_{j} \in[t_{1},\dots,t_{n}] \:\subset\:[0,T]$, typically the margin call dates or even at the coupon exchange dates.\\

For a matter of notations,  we define $\zeta_{j} = 1-z_{j}$ in order to better  distinguish between  the two regimes, but the control set is intended to be the same.\\
At this point, we need to remark - from the definitions of\emph{ contingent CSA} and (bilateral) \emph{CVA process}, 2.2.6 e 2.2.7 -   that the switching controls enter and affects the dynamic of this processes which are described by the system of stochastic differential equations assumed above. In fact, as we know, switching to full collateralization implies $CVA=0$ (and $BCVA=0$), that is $S_{t}=  S_{t}^{rf} = Coll^{perf}_{t}$. This means no counterparty risk and so the default intensity dynamic $d\lambda_{t}$ won't be relevant, just $dX$ will be considered until $z=0$, namely until the collateralization will be kept active ($\zeta =1$). So, formally, we have:\\

$if \: \big\{z_{j}= 1\big \} \: and \: \{\tau_{j}\leq t<\tau_{j+1}\}\Rightarrow \;$
\begin{eqnarray*}
  D_{t}^{C} &=& D_{t} \Rightarrow\\
  S_{t}^{C} &=& S_{t} \Rightarrow\\
  BCVA_{t}^{C} &=& BCVA_{t} \forall\:t\in[0,T\wedge \tau]
\end{eqnarray*}

so that the relevant dynamic to model the BCVA process in this regime is
\begin{eqnarray*}
  dX_{t} &=& \mu(t,X_{t})X(t) dt + \sigma(t,X_{t}) X(t) dW_{t}^{x};\quad X(0)=x_{0}\\
  d\lambda_{t} &=& \gamma(t,\lambda_{t})\lambda(t)dt + \nu(t,\lambda_{t}) \lambda(t)dW_{t}^{\lambda}; \quad \lambda(0)=\lambda_{0}
\end{eqnarray*}

$if \: \big\{z_{j}= 0 \big \} \; and \; \{\tau_{j}\leq t<\tau_{j+1}\}\Rightarrow $
\begin{eqnarray*}
  D_{t}^{C} &=& D_{t}^{rf} \Rightarrow\\
  S_{t}^{C} &=& S_{t}^{rf} =Coll_{t}^{Perf}  \Rightarrow\\
  BCVA_{t}^{C} &=& 0 \forall\:t\in[0,T\wedge \tau]
\end{eqnarray*}

so that the relevant dynamic  to model the process in this regime will be just
\begin{equation*}
  dX_{t} = \mu(t,X_{t})X(t) dt + \sigma(t,X_{t}) X(t) dW_{t}^{x};\quad X(0)=x_{0}
\end{equation*}


To ease the notation, we set our system dynamic in vectorial form defining\\
\begin{center}
$dY^{\mathcal{C}_{ad}}(t):=\left[
  \begin{array}{c}
     dt\\
    dX_{t} \\
    d\lambda_{t} \\
    dZ \\
  \end{array}
\right]$, $ \qquad Y^{\mathcal{C}_{ad}}(0)=\left[
  \begin{array}{c}
     t=0\\
     x_{0} \\
    \lambda_{0} \\
    Z_{0}=1 \\
  \end{array}
\right ]$
\end{center}
where we are left to define the set of \emph{admissible switching control} $\Big\{ \mathcal{C}_{ad} \subseteq \mathcal{C}=\big\{ \mathcal{T}, \mathcal{Z} \big\} \Big\}$, admissible in the sense that our stochastic (controlled) system described by $dY^{\mathcal{C}_{ad}}$ has a unique solution.\\

\subsection{ \textit{Modeling the objective functional.}}

The last element left to model to complete the picture is the objective functional of the agent, namely counteparty $A$.
In general, the formulation of the objective functional for a  switching control minimization problem can be stated as follows:
\begin{equation*}
    J^{(\mathcal{T},\mathcal{Z})}(y)=\inf_{ u  \in\{\mathcal{C}_{ad}\}} \mathbb{E}^{y} \bigg[ \int_{t=0}^{T} (\exp^{-r(T-s)}) F_{Z}(Y^{u}(s))ds + G(Y^{u}(T) )\mathbbm{1}_{\{T<\infty\}} + \sum_{\tau_{j} < T} K\big( Y^{u}(\tau_{j}), z_{j} \big) \bigg| \mathcal{F}_{t} \bigg]
\end{equation*}
where for notational convenience has been set $u=:\{\mathcal{T}, \mathcal{Z} \}$\footnote{So, $u$ contains $\mathcal{C}_{ad}$ the set of admissible controls in which the solution must be searched.} while the function $F(.)$ represents the agent objective that is in our case the running cost function, $G(T)$ is the so called ''\emph{reward function}'' and $K(.)$ the instantaneous \emph{switching costs} of the problem.\\
 As regards our problem, we have that:
\begin{description}
  \item[a)] For what concerns the running  cost function $F(.)$, we need to distinguish between the two switching regimes.\\ So if $\{z_{j}=1\}\;, ''zero\; collateral''$, we know from the latter section that the agent objective is to minimize the BCVA costs, in particular we assume (mainly for technical reasons) that  counterparty $A$ be interested  in the minimization of the BCVA variance respect to a given threshold $\delta$. Formally we have:
 \begin{equation*}
    F_{z}(Y^{u}(t))= z[BCVA(t)- \delta]^{2}  \Rightarrow \; ''\min \:BCVA \: variance''
 \end{equation*}
 where $z$ is the control switching indicator that multiply the BCVA because it is set to $z=1$ at inception and it can be zeroed when counterparty $A$ decide ''optimally'' to switch to full collateral, setting $z=0$.
 Here  the threshold $\delta$  can be seen as an exogenous parameter, for example it can be a given level of capital or liquidity allocated by the risk management function or by the treasury as a target of the problem.  An other possibility is to set it, coherently with the definition of variance,  as the expected BCVA value calculated at inception in $t=0$( alternatively one can also generalize the formulation by making the threshold endogenous and setting it as a control  of the problem possibly depending on time).\\

As concerns the other regime $\{z_{j}=0\}\;,''full \;collateral'' $, we recall from the definitions of past section that - being collateralization perfect/full - no partial, minimum transfer amounts and reuse/rehypotecation clauses are considered and we have assumed just cash or asset highly liquid as collateral  - we have shown (see 2.11 and 2.12) that this is equal to the NPV process. But we are going to show that switching to full collateral generates two possible expected costs to take in account in the objective functional.  Recalling also the existence of the same asset $B_{t}^{borr}$, $B_{t}^{opp} $, $B_{t}^{rem}$ defined in section 2.3 on funding and collateral\footnote{To ease notation we set to zero the basis spreads $bp_{t}$, $\bar{bp}_{t}$  over the risk free rate. } we have the following cases.
\begin{enumerate}
  \item If $\zeta_{j}=1$ and $NPV(s) >0$ $(s\geq\tau_{j})$, the counterparty (B) has to post the collateral in a segregated account that must be remunerated at the free risk $r$ (by the external funder)  but $A$ cannot use or invest that amounts so she has an opportunity costs, say $r_{opp}$ (set as a constant). Because this costs has to be considered until the collateral is active (next switching) and the NPV is positive, we can define the integral of the running costs function (which is taken until the next switching time $\tau_{j+1}$) as
      \begin{equation*}
        \int_{s}^{T\wedge\tau_{j+1}} \beta(v)^{+} [NPV(v)]^{+}ds = \int_{s}^{T\wedge\tau_{j+1}} \exp-(r_{opp} - r)v  [NPV(v)]^{+}ds \; s\in[\tau_{j},T].
      \end{equation*}
  \item By the other side, if $\zeta_{j}=1$ and $NPV(s) <0$ $(s\geq \tau_{j})$, A  has to post the collateral in a segregated account that must be remunerated at the free risk $r$ (by the external funder)  but $A$ now has to find the money to do this so she has a funding cost, say $r_{borr}$. Because this costs are alive until the collateral is active and the NPV is negative, we can define the integral of  running costs function (taken until the next switching time $\tau_{j+1}$) in this case as
      \begin{equation*}
         \int_{s}^{T\wedge\tau_{j+1}} \beta(v)^{-} [NPV(v)]^{-}ds  =\int_{s}^{T\wedge\tau_{j+1}}  \exp-(r_{borr} - r)v [NPV(v)]^{-}ds \; s\in[\tau_{j},T]
      \end{equation*}
\end{enumerate}
So we can put the things together and being
\begin{equation*}
   \int_{s}^{T\wedge\tau_{j+1}}  [NPV(v)]ds = \int_{s}^{T\wedge\tau_{j+1}}  [NPV(v)]^{-}ds + \int_{s}^{T\wedge\tau_{j+1}}  [NPV(v)]^{+}ds
\end{equation*}
and setting
\[
 R(s) = \begin{cases}
       - \beta(s)^{-}= -\exp-(r_{borr} - r)s  &,\; if \;\{\zeta_{j}=1\} \;and\; NPV<0\\
        \beta(s)^{+} = \exp-(r_{opp} - r)s &, if \; \;\{\zeta_{j}=1\} \;and \;NPV>0
\end{cases}
\]
and $0$ otherwise\footnote{Note we have set the minus before the exponential in the case in which the $NPV$ is negative in order to obtain a positive value, in fact  we've considered always positive values as costs according to the minimization objective.}, we get that the running switching cost function for the bilateral case is the following
\begin{eqnarray}
  F_{\zeta}(Y^{u}(s)) &=&  \bigg(    \int_{s}^{T\wedge\tau_{j+1}} \beta(s)^{+}  [NPV(v)]^{+}ds   + \int_{s}^{T\wedge\tau_{j+1}}  \beta(s)^{-}  [NPV(v)]^{-}ds  - NPV(s)\bigg) \nonumber\\
   &=&  \bigg(    \int_{s}^{T\wedge\tau_{j+1}}R(s)  [NPV(v)]ds - NPV(s) \bigg)  \; for\:s\in[\tau_{j},T],\:
\end{eqnarray}
$\forall \:\tau_{j},z_{j}    \in \{\mathcal{T}, \mathcal{Z}\} $.  Let us note  that before the integral we should set a summation over $j\leq M$ to consider all the possible switches but we set it later in the general objective functional; therefore we have set the minus before the $NPV(s)$ because if it's positive the counterparty $A$ receives the collateral which must be subtracted to the expected costs, if it's negative it must be added, being a sort of instantaneous cost or revenues (depending on the sign). In effects it can be incorporated in the instantaneous cost part of the functional too, but  this make the switching boundary stochastic for which the analysis in more involved and the optimal strategy may not exists\footnote{We refer to Hamadene \& Zhang (2010) for details.}.\\
So, because of our problem is a minimization, given the structure and the definition for  $ F_{\zeta}(Y^{u}(s))$ it can be positive or negative because of the last term ($NPV(s)$), we square it by keeping the same structure of the other running function related to the zero collateral case, so that we have
\begin{equation*}
  if \: \{z_{j}=0\} \rightarrow   F(Y^{u}(s))= \bigg(  \bigg(   \int_{s}^{T\wedge\tau_{j+1}}  R(s) [NPV(v)]ds - NPV(s) \bigg) - \delta \bigg)^{2} \; for\: s\in[\tau_{j},T]
\end{equation*}
and $\forall \:\tau_{j},z_{j}  \in \{\mathcal{T}, \mathcal{Z} \}$.
Here the threshold $\delta$ is set equal to the running cost function of the other regime switching, but it can be modeled differently.
  \item[b)] for what concerns the function $G(.)$ or \emph{reward function}, because of our problem is of finite time type and because of in $T$ - which is the problem terminal state/time - we can get or the terminal collateral value which equals (minus)\footnote{The minus is due to the minimization costs setting, so if $NPV{T}$ is positive it is not a cost, so we put the minus ahead.} the $NPV(T)$ if $z_{T}=0$ (or $\zeta_{T}=1$)\footnote{Note that here we are making an abuse of notation setting $z_{T}$ instead of $z_{M}$ but these are meant to be the same, that is $M=T$ for the reward function. Therefore, this is one of the  way to model this function that can be modeled in a more complicated way. }, or nothing, because the CVA is zero or almost zero, say $\epsilon$, for $t\rightarrow T$ if no collateralization is kept active $z_{T}=1$ (or $\zeta_{T}=1$).  Formally we set, as we work with the variance of the running costs function respect to a given threshold $\delta$, we have,
\begin{equation*}
    G(Y^{u}(T)) = (\epsilon-\delta)^{2} \times\mathbbm{1}_{\{z_{T}=1\}} + (-NPV(T)-\delta)^{2}\mathbbm{1}_{\{z_{T}=0\}}.
\end{equation*}
Actually, because it's possible in theory that the agent decide to switch just before $T$, to set off this type of behavior\footnote{That is typical of the continuous time framework, when we will tackle the implementation, in discrete times this won't be relevant anymore. } which causes a jump in the value function and to be more general and formal we set $z_{T^{-}}:=\lim_{t\rightarrow T}z_{t}=z_{T}$ (and similarly for $\zeta_{T^{-}}$),  and setting for convenience $NPV(T^{-})=NPV(T) $ and  $\epsilon=0$, we can rewrite the reward function, that is even our terminal/boundary condition, as
 \[  G(Y^{u}(T))=
  \begin{cases}
         (-NPV(T)\zeta_{T^{-}}- \delta)^{2}   \Rightarrow & \;,if \: collateral \:is\: active  \\
         (0 z_{T^{-}}-\delta)^{2}  \Rightarrow &,   \; otherwise\:\:no\: collateral
           \end{cases}
           \]
Here is important to remark that the valuation of the control problem is set in $t$ with $t<\tau\wedge T$ and all the value processes  are intended as pre-default values and clearly $\tau=\infty$ in $T$.

\item[c)] What is left to model is the \emph{instantaneous switching costs function }. It takes the fixed costs that counterparty has to consider when switches regime. A typical working formulation is the following:
    \begin{equation*}
        K\big( Y^{u}(\tau_{j}), z_{j} \big) = \sum_{j\geq1}^{M}e^{-r\tau_{j}} c_{z_{j}}(t)\mathbbm{1}_{\{ \tau_{j}<T\}}, \;\: \forall\:\:\tau_{j},z_{j}  \in \{\mathcal{T}, \mathcal{Z} \},
    \end{equation*}
   where the summation here replace the integral given that the instantaneous switching costs happen at the future optimal switching times. Here the  costs $c_{z_{j}}(t)$ are set as a deterministic $\mathcal{F}_{\tau_{j}}$-measurable and where the initial valuation  time is $t=0$ to ease the notation.
\end{description}

Putting all these pieces together we get  the \emph{general (bilateral) objective functional} for our problem
\begin{eqnarray}
    J^{(\mathcal{T},\mathcal{Z})}(y)&=& \inf_{ u  \in\{\mathcal{C}_{ad}\}} \mathbb{E}^{y} \Bigg[ \int_{t=0}^{T} (\exp^{-rs})\bigg( \sum_{j\geq 1}^{M}\big[(CVA(s)-DVA(s)) - \delta \big]^{2}\mathbbm{1}_{\{z_{j=1}\}} \nonumber  \\
   &+& \Big(  (\int_{u}^{T\wedge\tau_{j+1}} R(s) [NPV(v)]du - NPV(s)\big)- \delta \Big)^{2}\mathbbm{1}_{\{z_{j=0}\}}\bigg)ds  \nonumber   \\
   &+&   \sum_{j\geq 1}^{M} e^{-r\tau_{j}} c_{z_{j}}(t)\mathbbm{1}_{\{ \tau_{j}<T\}}+   \bigg(- NPV(T)\zeta_{T^{-}} - \delta\bigg)^{2} \bigg| \mathcal{F}_{t} \Bigg]\;\: \forall \:s,\tau_{j}\in[t,T\wedge \tau]
\end{eqnarray}
that can be represented in a more compact way, by setting
\[
 F_{Z}(y,u) = \begin{cases}
       \ \sum_{j}\big[(CVA(s)-DVA(s)) - \delta \big]^{2} + (0 z_{T^{-}}-\delta)^{2} & if \{z_{j}=1\}\\
       \sum_{j}\Big( \big( \int_{u}^{T\wedge\tau_{j+1}}  R(s) [NPV(v)]du - NPV(s)\big)- \delta \Big)^{2} + (-NPV(T)\zeta_{T^{-}}-\delta)^{2}  & ,\;if\; \{z_{j}=0\} .
        \end{cases}\]
$\forall \:s,\tau_{j}\in[t,T\wedge \tau]$, which allow us to write our control problem as follows
\begin{eqnarray}
    J^{(\mathcal{T},\mathcal{Z})}(y)=&=& \inf_{ u  \in\{\mathcal{C}_{ad}\}} \mathbb{E}^{y} \Bigg[ \int_{t=0}^{T} (\exp^{-rs}) [F_{Z}(y,u)]ds \nonumber
      +    \sum_{j\geq 1}^{M} \exp^{-r\tau_{j}}c_{z_{j}}(t)\mathbbm{1}_{\{ \tau_{j}<T\}} \bigg| \mathcal{F}_{t} \Bigg]
\end{eqnarray}

subjecting everything to the given dynamics $dY$ and the admissible control set $\mathcal{C}_{ad}$ it tells us that the objective is to find , if it exists and is unique, the optimal sequence of switching times and indicators $(z_{j}, \tau_{j})^{*}=u^{*}$ that minimize the expected costs functional, that is formally to find the \emph{value function} of the problem $V(y)$ and $u^{*} \in \mathcal{C}_{ad} $ such that
\begin{equation}
    V(t, x, \lambda;u) := V(y)= \inf \big\{ J^{u}(y) ; u \in \mathcal{C}_{ad} \big\} = J^{u^{*}}(y)
\end{equation}
with the \emph{terminal condition} in $T$ on the value function that will be\\
\begin{equation}
    V(T, x ;u) := \inf_{u}\{(-NPV(T)\zeta_{T^{-}}-\delta)^{2}, (0  z_{T^{-}}- \delta)^{2}\}.
\end{equation}

\subsection{\textit{ Solution existence and uniqueness.}}
 	
The switching control problem (3.5-3.7) formulated in the latter section can be seen as a special two regime switching control problem with two different (running) cost functions and different dynamics for each regime. This type of problem admits both an analytical representation through (quasi) \emph{variational inequalities} with interconnected obstacles and a stochastic representation through  \emph{reflected backward stochastic differential equations } which are deeply connected (see Mottola (2013) for a detailed discussion of the issue).

The  main issues related to the solution of our problem can be sum up as follows:
\begin{description}
  \item[a)] \emph{the impossibility to assume (and guess) a  smooth value functions  $V^{i}$ for the "switching nature" of the problem and the complexity of the system of PDE which are actually partial-integro differential equations with interconnected obstcles ;}
  \item[b)] \emph{the non linearity of the cost functions $F^{i}$ (that do not grow linearly and  are not lipschitzian) and their recursive nature depending over time on the controls (the switching times $\tau_{j}$ and indicators $z_{j}$)};
\end{description}

But thanks to results due to Djehiche, Hamadene \& Popier (2008) on stochastic switching control problems (with multiple regimes in finite horizon), we are able to tackle - under fairly general assumptions -  the solution existence for our problem  via a stochastic approach based on generalized \emph{Snell envelope} tool.\\
In particular, we show  that the solution of the switching problem - namely the determination of the optimal switching strategy - coincides with the proof of the existence - via a \emph{verification theorem} - of the vector process $(Y^{z}, Y^{\zeta})$ solution of a system of Snell envelopes, where $Y^{z}$ (resp. $Y^{\zeta})$ indicates the optimal expected costs when collateral is set to zero (resp. one).\\

\textbf{Theorem 4.2.2 (Verification theorem).} \emph{Suppose there exist two processes $Y^{z}:= (Y^{z}_{t})_{0\leq t\leq T}$ and $Y^{\zeta}:= (Y^{\zeta}_{t})_{0\leq t\leq T}$ belonging to $\mathcal{K}^{p}$ such that for any $t\in[0,T]$ and a generic switching time $\tau:=\tau_{j}$
\begin{equation}
    Y^{z}_{t}=ess\: sup_{\tau\geq t }\mathbb{E}\bigg[\int_{t}^{\tau} e^{-r(\tau-s)}F^{z}(s,\mathcal{Y}_{s})ds + (Y^{\zeta}_{\tau} - e^{-r\tau}c^{z})\mathbbm{1}_{\{\tau <T\}} |\mathcal{F}_{t}\bigg], \; Y^{z}_{T}=0,
\end{equation}
\begin{equation}
    Y^{\zeta}_{t}=ess\: sup_{\tau\geq t }\mathbb{E}\bigg[\int_{t}^{\tau} e^{-r(\tau-s)}F^{\zeta}(s,\mathcal{Y}_{s}')ds + (Y^{z}_{\tau} - e^{-r\tau}c^{\zeta})\mathbbm{1}_{\{\tau <T\}} |\mathcal{F}_{t} \bigg], \; Y^{\zeta}_{T}=0.
\end{equation}
Then $Y^{z}$ and $Y^{\zeta}$ are unique. Therefore, the following is verified
 \begin{equation}
    (a) \;\; Y_{0}^{z}  = V(u^{*}) = \sup_{u\in \mathcal{C}^{ad}} (-J(y,u)).
 \end{equation}
(b) Defining the stopping times sequence $(\tau_{j})_{j\geq 1}$ as
 \begin{eqnarray}
   \tau_{1} &=& D_{0}(Y^{z} = Y^{\zeta} - c^{z}) \; \;on \; \{\tau_{1}< T \} \; (and \: \{ z_{0}=1\}); \\
\tau_{2} &=& D_{1}(Y^{\zeta} = Y^{z} - c^{\zeta}) \; \;on \; \{\tau_{2}< T \} \; (and \: \{ z_{0}=1\});
 \end{eqnarray}
and in general for $j\geq 1$
\begin{eqnarray}
   \tau_{2j+1} &=& D_{\tau_{2j}}(Y^{z} = Y^{\zeta} - c^{z}) \; \; \;on \; \{\tau_{2j+1}< T \}  \\
\tau_{2j+2} &=& D_{\tau_{2j+1}}(Y^{\zeta} = Y^{z} - c^{\zeta}) \; \; on \; \{\tau_{2j+2}< T \} .
 \end{eqnarray}
Then the switching control strategy $u^{*}:=((\tau_{j}^{*})_{j\geq1},(Z_{j})_{j\geq 1}^{*}) $ is optimal.}\\

\emph{Proof.} For the proof we refer to  Mottola (2013) which follows the same lines of the one in the paper of Djehiche \emph{et al.} (2008).\\

For the proof of uniqueness and of the deep result that connects the Snell envelope vector solution  $(Y^{z}, Y^{\zeta})$ with the solution of \emph{backward SDE with reflection} which are shown to be (under some assumptions) also \emph{viscosity solutions} of  system of \emph{PDE with interconnected obstacles},  we also refer to Mottola (2013).

\section{\large Numerical Solution: program description and results}

\subsection{  \textit{Numerical procedure definition}}

In this last section we build on the theoretical results and insights of the past sections to define a suitable numerical procedure in order to solve our switching problem. 
In particular, we recast the problem as an \emph{iterative optimal stopping time} whose main ingredients for the numerical solution are a Monte Carlo generator for the paths simulation, a backward induction procedure founded on the \emph{dynamic programming principle } and an approximation method (via \emph{least-square regression}) of the nested conditional expectation at each step of the recursion. A detailed description of the procedure implemented can be found in Mottola (2013).\\ 

\subsection{  \textit{Numerical procedure definition and analysis}}



Firstly, let us recall the main ingredients and assumptions of the problem we want to tackle numerically. Firstly - following Carmona \& Ludkovski (2006) - we  recast our problem as an iterative optimal stopping problem, that is
\begin{eqnarray}
 V^{0}(t,y,Z) &:=& \mathbb{E}\bigg[ \int_{t}^{T} e^{-r(T-s)}F^{Z}(s,\mathcal{Y}_{s})ds\big| \mathcal{Y}_{t}=y \bigg] ,\dots, \\
V^{j}(t,y,Z) &:=& sup_{\tau\in \mathcal{T}}\mathbb{E}\bigg[ \int_{t}^{T\wedge \tau} e^{-r(T-s)}F^{Z}(s,\mathcal{Y}_{s})ds + SW^{j,Z}(\tau,\mathcal{Y}_{\tau})\big| \mathcal{Y}_{t}=y \bigg]
\end{eqnarray}
where $t\in[0,T]$, $\tau:=\tau_{j}$ (here we make an abuse of notation, but is intended that we assume similar conditions on default times $\tau$ and switching times as in section four)  and $j=0,1,2,\dots,M$ represents a given number of switches allowed, $Z=\{z, \zeta\}$ the switching indicators, $y$ indicates the realization of our markovian vector dynamic $\mathcal{Y}_{t} = (X_{t}, \lambda_{t})  $ set in section 3.1) intended to be defined under the real/objective measure $\mathbb{Q}$  - we remind that the value function have to be intended always negative with a minus ahead, to give sense to the minimization problem and that the allowed switching time are finite because of the presence of switching costs - while $SW(.)$ is the \emph{switching/intervention operator} which guide the recursion and takes the following expression
\begin{eqnarray*}
   SW^{l,Z=z}(\tau,\mathcal{Y}_{\tau})&:=& \max_{Z \in\mathcal{Z}} \{V^{l-1}(\tau,y,Z=\zeta)- c_{\tau}^{Z=z}\} \\
   &=&  \{V^{l-1}(\tau,y,Z=\zeta)- c_{\tau}^{Z=z}\},
\end{eqnarray*}
given that the maximum is trivial in our case with only two regimes $\{z,\zeta\}$.
For what concerns our modeling choice for the stochastic vector dynamic, because of we need numerous simulation of the exposures of the given contract (represented by an IRS) not only at the valuation date $t=0$  but also at forward times and given that the exposures depend - for what concerns the interest rate modeling part -   on the libor/forward rate evolution, a \emph{libor market model } dynamic would be the natural modeling choice. But given the complexity of the problem and the high number of simulation and valuation date, using a LMM dynamic would be a cumbersome numerical task.\\
So, without great loss of generality, the choice has been in favor of a more manageable short rate model. In order to  ensure a grater variety of shapes of the simulated term structure  we use for the computations a \emph{shifted two factor gaussian }short rate model (say G2++). Indeed, the dynamics for $X$ becomes
\begin{equation*}
    X_{t}:= r_{t} = y_{t}+z_{t}+\varphi_{t}, \;\;r(0)=r_{0}
\end{equation*}
where the processes $\{y_{t}: t\geq0\}$ and $\{z_{t}: t\geq0\}$ satisfy
\begin{eqnarray*}
  dy_{t} &=& -\mu y_{t}dt+ \sigma dW_{t}^{1}, \;\; y(0)= y_{0}\\
  dz_{t} &=& -\nu y_{t}dt+ \eta dW_{t}^{2}, \;\; z(0)= z_{0}\\
  d\langle y,z\rangle_{t} &=& d\langle W^{1}, W^{2}\rangle_{t} = \rho_{y,z}dt
\end{eqnarray*}
where $(W^{1}, W^{2})$ is a two-dimensional $\mathbb{Q}$ Brownian motion with instantaneous
correlation $\rho \in[-1,1] $, $\varphi_{t}$ is  a deterministic function that ensures the fitting on the initial market term structure and $\:\{r_{0}, \mu,\nu,\sigma,\eta\}$ are positive constant parameters representing the drifts and instantaneous volatilities of $r_{t}=X_{t}$ (here to ease the notation we omit to highlight the market price of risk in the drift dynamic).
Because of diffusion parameters do not change under change of measure and  we are not under the pricing measure we  have chosen to calibrate it on the historical volatility of the reference rate underlying the contract, typically the \emph{six months euribor}. The remaining parameters are instead determined by standard calibration on the cap-floor quotes.\\
For what concerns the model for the default intensity $\lambda_{t}$, the choice has been in favor of a typical \emph{Cox process} with stochastic intensity $\lambda$, assumed $\mathcal{F}_{t}$-adapted, positive and right continuous ($\mathcal{F}_{t}^{\lambda } = \sigma(\{\lambda_{s} : s\leq t\}))$ having dynamic of CIR type given below
\begin{eqnarray*}
 d\lambda_{t}   &=& \kappa(\gamma - \lambda_{t})dt + \upsilon \sqrt{\lambda_{t}}dW_{t}^{\lambda}  \\
   \lambda(0)&=& \lambda_{0}
\end{eqnarray*}

where $W_{t}^{\lambda}$ is also a standard $\mathbb{Q}$-Brownian motion. We also recall that the \emph{cumulated intensity} or \emph{hazard process} is defined as $\Lambda(T)  = \int_{0}^{T}\lambda_{t}dt $ which is a random variable (absolutely) continuous and increasing. In this framework $\Lambda(T) $ play a central role for jump time $\tau$ simulation given that the first jump time of the process, transformed through its cumulated intensity, is an exponential random $\xi$ variable independent of $\mathcal{F}_{t}$ (being part of the defaultable filtration $\mathcal{H}_{t}$), that is
\begin{equation*}
    \Lambda(\tau)  =\xi \; \Longrightarrow \:(inverting)
\end{equation*}
\begin{equation*}
    \tau  :=\Lambda^{-1}(\xi)
\end{equation*}
The procedure to simulate the default times $\tau$ is the standard one (see for example Brigo \& Mercurio (2006)).
The vector process parameters $(\kappa, \gamma, \upsilon)$ of positive constants, can be in general calibrated to credit/bond market spreads or CDS spreads of specific names. In particular, we will distinguish between two different cases using the market quotes of two different names (labeled as LOW and HIGH counterparty risk)  for volatilities and drift of the default intensities in order to investigate the effects on the value function and the optimal switching strategy of the problem.\\
In the following tables we summarize the SDEs parameters and those related to the perfect collateralization regime and funding, the borrowing $r_{borr}$, opportunity costs $r_{opp}$ and for the switching costs $c_{Z}$ that we need to set in computations.  The parameter $\delta$ is set equal to zero and to meaningful value in order to check the impact on the solution and on the optimal switching strategy. Here follows a table of the variables and the related parameters  of our model.
\begin{center}
\begin{tabular}{|c|c|c|}
\multicolumn{3}{}{ } \\
\hline
$X_{t}$ &  $\lambda$ & Other parameters\\
\hline
  $\mu$ & $\kappa$ &$r_{opp}$ \\
  $\nu$ & $\gamma$ &$r_{borr}$ \\
  $\sigma$ & $\upsilon$ &$R_{c}$ \\
  $\eta$ &            &$c_{z}$\\
  $\rho$ &            &$c_{zeta}$\\
         &            &  $\delta$\\
\hline
\end{tabular}
\end{center}


The steps of the path discretization  and of the chosen implementation procedure - done under the symmetry hypothesis\footnote{See proposition 5.2.1 on symmetry of Mottola (2013)} - are detailed in Mottola (2013).\\

Here, let us just explicit the program the we need to implement for the solution of our  switching control problem.\\
\begin{description}
  \item[a)] Let us consider for simplicity  two calculation times $t_{1}$ and $t_{2}$, thanks to the stated optimality principle, the counterparty has to decide between immediate switch at $t_{1}$ to the other regime $z$ (or $\zeta$) versus no switching and wait until $t_{2}$. So, discretizing the recursive equations (4.1)-(4.2) for the value function, we run the following program:
        \begin{eqnarray*}
          V^{l}(t_{1},\mathcal{Y}_{t_{1}},Z) &=& \max\bigg( \mathbb{E}\big[\int_{t_{1}}^{t_{2}} F^{Z}(s,\mathcal{Y}_{s})ds + V^{l}(t_{2}, \mathcal{Y}_{t_{2}},Z)|\mathcal{F}_{t_{1}}\big], \: SW^{l,Z}(t_{1},\mathcal{Y}_{t_{1}})\bigg)  \\
           &\simeq& \max\bigg( F^{Z}(t_{1},\mathcal{Y}_{t_{1}})\Delta t +  \mathbb{E}\big[  V^{l}(t_{2}, \mathcal{Y}_{t_{2}},Z)|\mathcal{F}_{t_{1}}\big], \\
           & & \{V^{l-1}(t_{1},\mathcal{Y}_{t_{1}},Z=\zeta)- c_{t_{1}}^{Z=z}\}\bigg)
        \end{eqnarray*}
    formulation that becomes in our cost minimization problem as follow
     \begin{eqnarray*}
          V^{l}(t_{1},\mathcal{Y}_{t_{1}},Z)  &\simeq& \min\bigg( F^{Z}(t_{1},\mathcal{Y}_{t_{1}})\Delta t +  \mathbb{E}\big[  V^{l}(t_{2}, \mathcal{Y}_{t_{2}},Z)|\mathcal{F}_{t_{1}}\big],  \{V^{l-1}(t_{1},\mathcal{Y}_{t_{1}},Z=\zeta) + c_{t_{1}}^{Z=z}\}\bigg)
        \end{eqnarray*}
    where here $BCVA^{\Delta}_{t_{1}} = \mathcal{Y}_{t_{1},z}$ (and similarly $Coll^{c}_{t_{1}} = \mathcal{Y}_{t_{1},\zeta}$), so that the last one can be made more explicit as follows\\

        $\Big\{if \: in\:t_{2}\: \{Z=z\} \Big\}$ $\Longrightarrow$
    \begin{equation}
          V^{l}(t_{1},\mathcal{Y}_{t_{1}},Z) \simeq \min\bigg( (BCVA^{\Delta}_{t_{1}}-\delta)^{2} \Delta t +  \mathbb{E}\big[  V^{l}(t_{2}, \mathcal{Y}_{t_{2}},z)|\mathcal{F}_{t_{1}}\big],  \{V^{l-1}(t_{1},\mathcal{Y}_{t_{1}},\zeta) + c_{t_{1}}^{z}\}\bigg)\;\\
    \end{equation}

    $\Big\{if \: in\:t_{2}\: \{Z=\zeta\} \Big\}$ $\Longrightarrow$
    \begin{equation}
          V^{l}(t_{1},\mathcal{Y}_{t_{1}},Z)  \simeq \min\bigg( (Coll^{c}_{t_{1}}-\delta)^{2} \Delta t +  \mathbb{E}\big[  V^{l}(t_{2}, \mathcal{Y}_{t_{2}},\zeta)|\mathcal{F}_{t_{1}}\big],  \{V^{l-1}(t_{1},\mathcal{Y}_{t_{1}},z) + c_{t_{1}}^{\zeta}\}\bigg).
  \end{equation}
         Here the conditional expectation of the cost function $\mathbb{E}[.|\mathcal{F}_{t_{1}}]$ represents the value of waiting - like the well known \emph{holding/continuation value} -  until the next switching time remaining in the same regime.
         At this point, to calculate this conditional expectation we have chosen to reapply the  Longstaff-Schwartz algorithm to the simulated paths ($BCVA^{\Delta}$ and $Coll^{c}$)  that enter  our cost functions which we recall  
         here (discretized in the case of an interest rate swap)
    \begin{eqnarray*}
    BCVA^{\Delta}(t_{i}) &=& (1-R_{c})\mathbb{E} \bigg[NPV(t_{i+1})\big| \mathcal{F}_{t_{i}}\bigg]\Delta t \lambda(t_{i},t_{i+1}) + BCVA(t_{i+1})  \\
    &=&  (1-R_{c})\bigg[\sum_{s} B_{t_{i}}\xi_{t_{i}}(L(t_{i},t_{i+s})-k)\bigg](t_{i+1}-t_{i}) \lambda(t_{i},t_{i+1})+ BCVA(t_{i+1})\;\;\nonumber
    \end{eqnarray*}
    \begin{eqnarray*}
    Coll^{c}(t_{i}) &=& R_{fund}(t_{i+1}-t_{i})\mathbb{E} \bigg[NPV(t_{i+1})\big| \mathcal{F}_{t_{i}}\bigg]\Delta t  + NPV(t_{i+1})  \\
    &=&  R_{fund}(t_{i+1}-t_{i})\bigg[\sum_{s} B_{t_{i}}\xi_{t_{i}}(L(t_{i},t_{i+s})-k)\bigg](t_{i+1}-t_{i}) + Coll^{c}(t_{i+1})\;\;\nonumber
    \end{eqnarray*}
 $\forall t_{i} \in [0,T]$ and paths $\omega$, with $\{t_{i} \leq \tau\}$ and $\{\tau_{j}\leq t_{i} < \tau_{j+1} \}$ (this  means that the \emph{default time} can happen and is bucketed at the valuation time $t_{i}$ which can coincide with a switching time $\tau_{j}$).

Once this is done,  the iterative procedure can be started firstly initializing the value function at the maturity setting $V^{l}(T, \mathcal{Y}_{T},Z) = 0$ $\forall\: l,\:z $ and then going backward computing iteratively the \emph{continuation values} $\mathbb{E}\big[  V^{l}(t_{n}, \mathcal{Y}_{t_{n}},z)|\mathcal{F}_{t_{n-1}}\big]$ along each paths. Then adding the term $ F^{Z}(t_{n},\mathcal{Y}_{t_{n}})\Delta t $  and taking the minimum as in equation 4.3 (or 4.4), averaging and discounting back to inception $t= 0$ one obtains the value function for the problem (actually the value functions are two as shown in section 4 depending on the initial regime $Z$ from which starts the iterative optimal stopping program).
By keeping trace of the paths in which is optimal to switch/not switch one can easily recover the optimal switching strategy $\{\tau_{j}^{*}, Z_{j}^{*}\}_{j\geq 1}$.\\


  \item[b)] In addition or alternatively to the procedure shown above,  one can keep trace of the \emph{smallest optimal switching time} $\tau^{l}(t_{n},\mathcal{Y}_{t_{n}}, Z ) $ that is also computed in backward manner and  represents for every path \emph{the smallest time in which is optimal to switch to an other regime}. The formal (recursive) definition is the following:
        \[ \tau^{l}(t_{n},\mathcal{Y}_{t_{n}}, Z ) =
  \begin{cases}
        \tau^{l}(t_{n+1},\mathcal{Y}_{t_{n+1}}, Z ), &$\;$ no switch \\
        n, &  $\;$ switch
        \end{cases}
\]
and the \emph{value function} will be given by the sum path by path of the future costs (discounted) until the optimal switching times, namely
 \begin{equation}
          V^{l}(t_{n\Delta t},\mathcal{Y}_{t_{n \Delta t}},Z)  = \mathbb{E}\Big[ \sum_{j=n}^{\tau^{l}} F^{Z}(j\Delta t, \mathcal{Y}_{j\Delta t})\Delta t + SW^{l,Z} (\tau^{l}\Delta t, \mathcal{Y}_{\tau^{l}\Delta t})|\mathcal{F}_{t_{n \Delta t}}\Big].
 \end{equation}
We conclude by noting that  the last formulation for the value function allows faster computations because here the regressions are not stored, they are just  used to update the switching times.

\item[c)]  The last thing that we need to highlight is how to recover the \emph{optimal switching boundary } for our problem. It represents the graph $(t,\mathcal{Y}_{t})$ of the boundaries at which our optimal switching strategy changes regime (for each $l,Z$). Keeping track of the minimal switching times $\tau^{l}$, one can reconstruct the switching boundary by drawing, at the end of the algorithm,   the graph of $\tau^{l}(0,\mathcal{Y}_{0}, Z )$ against $\mathcal{Y}_{t}$. Formally, the set
    \begin{equation}
        \big\{\mathcal{Y}_{n \Delta t} :\:  \forall \:n \:| \: \tau^{l}(0,\mathcal{Y}_{n}, Z ) = n  \big\}
           \end{equation}
    is defined as the empirical region of switching from a given regime $Z$ at time $n \Delta t$. It is important to recover the switching boundaries especially in a risk management view, as one can construct and check the optimal switching policies by simulating paths of $\mathcal{Y}$ and verifying the minor costs and benefits obtained by the implementation of the switching policy (also respect to the one that would  have been implemented without the flexibility of the switching type mechanism).
\end{description}

\subsection{ \textit{Numerical implementation and examples.}}

In this section we pass to show and analyze the main results of the implementation of the algorithm above defined for the solution of our stochastic switching control problem. From the results of sections 3, we are sure of the existence of a solution for our problem and also of the optimal switching strategy, although this strategy can also reveal to be of ''\emph{banal type}'' (as stated in  \emph{Definition 4.5.2} of Mottola (2013)) .\\
More specifically, in addition to the switching costs whose level can determine the convenience to switch, we expect that,  the volatility of the stochastic factors of our model, in particular the volatility and drift parameters of the default intensities would have a relevant impact on the switching strategy of the counterparty  that is brought to switch more often\footnote{In our model the funding costs are modeled as deterministic; of course in reality a counterparty which sees its default intensities increase see also the credit worthiness get lower with a likely increase also in the cost of funding. In this sense, one should need a stochastic  model also for funding costs correlated with default intensities or eventually including also \emph{liquidity spreads } issues. Here we work with a simpler deterministic hypothesis, which is enough to get some results, but with our numerical approach MC based it would not be difficult to generalize it adding an other stochastic factor (of course one should be prepared to deal with the increasing computational costs)}.
\begin{center}
\begin{tabular}{|c|c|c|c|}
  \hline
  \multicolumn{4}{|c|}{Market spot rates and Zero Coupon }\\
  \hline
   Maturity & Id &Spot Rates &Discount Factors\\
  \hline
  $1m$ &  EUR030 & $0.382\%$ & $0.9997 $\\
   $3m$ & EUR090 & $0.662\%$ & $0.9983 $\\
   $6m$ & EUR180 &$0.939\%$  & $0.9953 $\\
   $1y$ & EUR360 &$1.226\%$  & $0.9879 $\\
   $2y$ & EURS02 &$0.876\%$  &  $0.9827 $\\
   $3y$ & EURS03 &$0.985\%$  & $0.9710 $\\
  $4y $ & EURS04 & $1.151\%$ & $0.9553 $\\
  $5y$ & EURS05 &$1.331\%$  &   $0.9360 $\\
  $7y$ & EURS07 &$1.632\%$  &  $0.8929 $\\
  $10y$ &EURS10 &$1.949\%$ & $0.8245 $\\
  $12y$ &EURS12 &$2.089\%$ & $0.7802 $\\
  $15y$ &EURS15 &$2.204\%$ & $0.7211 $\\
  $20y$ &EURS20 &$2.211\%$ & $0.6457 $\\
  $25y$ &EURS25 &$2.202\%$ &$0.5802 $\\
  $30y$ &EUR.S30   &$2.193\%$&$0.5217 $\\
  \hline
\end{tabular}
\end{center}

In the following we focus the analysis on these variables and we report the results of the implementation of the program defined in section 4.2 applied to a simple \emph{defaultable interest rate swap, EURIBOR6m vs fixed rate} (for convenience the setting of the rate is in arrears) traded at par and maturing in one year $T=1Y$ from the inception $t$. The data used in computations  are taken at the reference date of 2012/06/15 from the main international data provider (\emph{Bloomberg} and \emph{Reuters}).\\
 Here we report the \emph{market yield curve} build on monetary rates (EURIBOR RATES) before the year and on the swap rates after the year,   and the ATM (flat) implied Cap volatilities

\begin{center}
\begin{tabular}{|c|c|}
  \multicolumn{2}{}{} \\
  \hline
  Maturity & Market implied vol  \\
  \hline
  $1Y$ &   $94.2\% $\\
$1.5Y$&  $93\%   $\\
$2Y$ &   $90.6\% $\\
$3Y$ &   $67.7\% $\\
$4Y$ &   $63.7\% $\\
$5Y$ &   $61.2\% $\\
$6Y$ &   $59.1\% $\\
$7Y$ &   $57.5\% $\\
$8Y$ &   $56.2\% $\\
$9Y$ &   $54.8\% $\\
$10Y$ &  $53.6\% $\\
$12Y$ &  $51.6\% $\\
$15Y$ &  $49.6\% $\\
$20Y$&   $48.8\% $\\
$25Y$&   $47.7\% $\\
 $30Y$&  $46.4\% $\\
\hline
\end{tabular}
\end{center}
Being traded at par, the strike rate used is the one year par swap rate $S01=0.91\%$. In order to keep a good representation of the reality by the model and to deal also with computational costs, the discretization step  used to simulate the vector dynamics and to update the switching strategy, has been chosen to be daily, that is $\Delta t =\frac{1}{N} $ where  $N=252$ (counting only the working days). The parameters of the $G2++$ process for $X_{t}$ calibrated on market implied (Black) volatilities, are the following
\begin{center}
\begin{tabular}{|c|c|}
  \multicolumn{2}{}{} \\
  \hline
  Calibrated parameters & Values  \\
  \hline
  $\mu$ &  $0.00013$ \\
  $\nu$ & $0.06730$\\
  $\sigma$ &$0.12924$\\
  $\eta$ & $0.14014$\\
  $\rho$ & $-0.99948$ \\
\hline
\end{tabular}
\end{center}
while the daily \emph{historical volatility } of the $EURIBOR6m$, calculated over the last year data sample, is $\sigma_{hist} = 0.12654$.
The calibration procedure is a standard minimization of sum of the square percentage difference between the model and the market price\footnote{Namely, we solve the following problem:
 \begin{equation*}
    \min_{\theta} \sum_{i=1}^{n_{payments}} \Big|\Big|  \frac{P^{mkt}(t,i) - P^{model}(t, i,\theta) }{P^{mkt}(t,i)} \Big|\Big|^{2}.
 \end{equation*}}.
Excluding the \emph{default intensities } on which we have decided to focus the analysis of the value function behavior, in calculations we have set the remaining parameters as follows:
\begin{itemize}
 \item the\emph{ risk-free rate }$ r_{free}$ being near to zero, if one considers as reference the 1Y yield from German bond, is set simply as constant $r_{free} = 0$;
  \item the\emph{ borrowing rate } $  r_{borr}$ should consider the particular balance-sheet and risk condition of the counterparty. For convenience we  assume it as  constant set to $r_{borr}=0.01$;
  \item similar considerations are valid for the\emph{ opportunity rate } $r_{opp}$ which is set equal to $r_{opp}=0.03$ by considering on average the rate of return of no risk free investments in the current stagnant market situation;
  \item the instantaneous switching costs $c_{z}$ and $c_{\zeta}$ are set as constants not greater than the $2\%$ of the notional of the underlying contract. Anyway, we show their impact on the switching strategy by varying their values;
  \item the recovery rate parameter $R_{c}$ is assumed fixed and equal to $40\%$ loss recovery if counterparty defaults;
   \item    the  parameter $\delta$ which enters in the cost functions is also set to zero (this means that counterparty would be adverse to every variation from zero of the running costs)  $\delta = 0$ and then varied.\\
\end{itemize}
As already mentioned, for what concerns the default intensities parameters we have calibrated  the model parameters $(\kappa,\gamma,\lambda_{0})$ on the  following CDS spreads quotes of two bank names (at the same date 15/6/2012)
\begin{center}
\begin{tabular}{|c|c|c|}
  \hline
  \multicolumn{3}{|c|}{CDS spreads: Dexia=HIGH, DB= LOW} \\
  \hline
  Bucket& Dexia &  DB \\
  \hline
  1Y & 915.716 & 93.648\\
  3Y & 817.403 & 156.834\\
  5Y & 782.0214 & 196.917\\
  7Y & 825.052 & 206.326\\
  10Y & 796.494 & 214.172\\
\hline
\end{tabular}
\end{center}

given the intention to investigate the results in case of HIGH (Dexia Subordinated)  and LOW (Deutsche Bank (Senior), DB)  default risk level for the counterparty. The calibration results are reported in the following table with the volatility parameter $\upsilon$  which is set taking the (daily)  CDS historical  volatility over the last two year.
\begin{center}
\begin{tabular}{|c|c|c|}
 \hline
  \multicolumn{3}{|c|}{$\lambda_{t}$ parametrization} \\
  \hline
  level & LOW  & HIGH\\
  \hline
  $\kappa$ & $1.03921$  &$0.30821$\\
  $\gamma$ &  $0.02120$ &$0.11220$\\
  $\upsilon$ & $0.20122$ &$0.44214$\\
  $\lambda_{0}$ & $0.04031$ &$0.20316$\\
  \hline
\end{tabular}
\end{center}
\begin{description}
  \item[\textbf{Results in "HIGH CASE"}.] From the application of the algorithm described in the last section and with the parametrization set as follows:\\
       - $\lambda_{t} = $ "HIGH";\\
       - $c_{z} =0 $ and $c_{\zeta} = 0$;\\
       - $\delta = 0$;\\
       - $Z=\zeta =0 $ in $T$, at maturity;\\
       - $N_{paths} = 1000$ which indicates the number of simulated paths; these are enough to derive the optimal switching strategy and to understand the value function behavior. In fact the problem has been tackled from a risk management view so that from a pricing view a greater number of paths should be run (other than a change of measure under which make calculations). \\
   Given the other parameters as defined above, we get the following main results.\\
   \begin{figure}[h!]
  \centering
  \includegraphics[  scale=0.55]{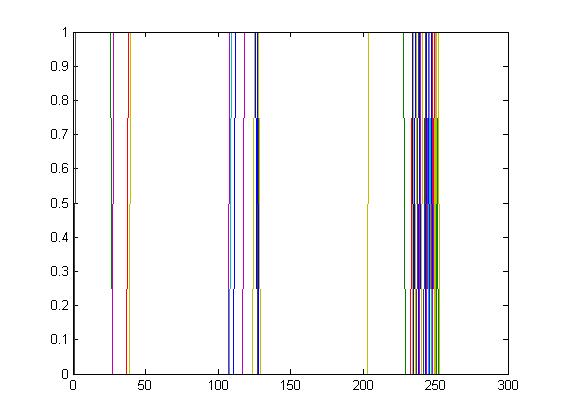}
  \caption{\emph{Graph of the switching indicators $Z=\{0,1\}$ over time $t\in[0,252]$} }\label{b}
  \end{figure}
   In Fig. 2 we have plotted the switching indicators $Z=\{0,1\}$ that are determined optimally by the application of the iterative switching algorithm (equations 4.3 and 4.4) over the time (discretized) domain $t \in [0,252]$. It shows almost four  region in which the switching over the paths and time have a daily frequency but also a big mass given the high number of paths in which results to be optimal to "switch and switch back" often. This was expected given the higher volatility of the default intensities and the fact that the instantaneous switching costs $c_{Z}$ are set to zero.
   \begin{figure}[h!]
  \centering
  \includegraphics[  scale=0.8]{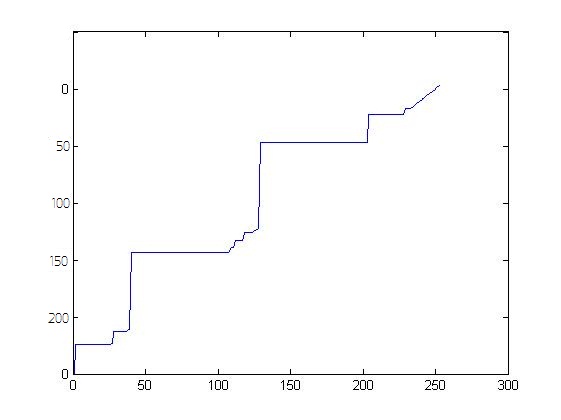}
  \caption{\emph{Graph of the minimum number of remaining switches for all the paths $N_{paths}$ over time $t\in[0,252]$} }\label{b}
  \end{figure}
  This behavior can be seen also from Fig. 3 in which we have plotted the minimum number of remaining (optimal) switches registered over the paths and time (as time elapses). Being the total number of switches equal to the number of time steps (253), by time elapsing also the number of switches decreases and the plot clearly shows in correspondence of flat lines and (negative) jumps the passage from of a "\emph{period of non activity}" in terms of switching to a "\emph{period of activity}".\\
   \begin{figure}[h!]
  \centering
  \includegraphics[  scale=0.55]{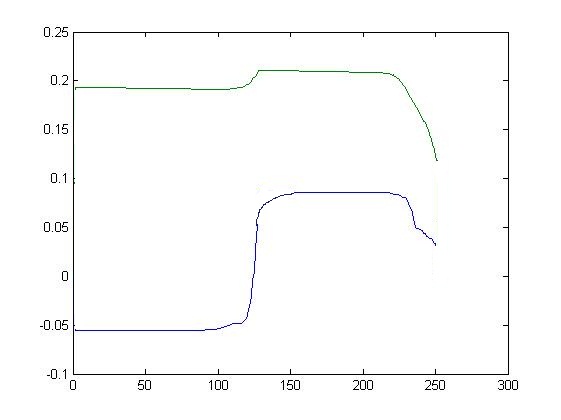}
  \caption{\emph{Graph of the switching boundaries as defined in equation 5.9 ($t\in[1,250])$} }\label{b}
  \end{figure}
We also report in Fig. 4 the graph of the optimal switching boundaries for our problem as defined in equation 4.5. The upper line shows the optimal values of the paths of $\mathcal{Y}^{z}$ above which it becomes optimal to switch, while the lower line indicates the paths for $\mathcal{Y}^{\zeta}$ above which is not optimal to switch. So the area between the two boundaries can be defined as the ''\emph{non action/switching area}''.\\
This boundaries are central in the risk management view, given that by confronting these values with that of simulated paths of the stochastic vector dynamic that guide  a given underlying claim one can study and define an optimal strategy that maximizes the hedging portfolio wealth or minimizes the related costs.\\
In the following table we summarize the results of the value function (with initial condition $\zeta =0$) for our problem, with a notional (of the underlying contract) $Notional = 1000$
\begin{center}
\begin{tabular}{|c|c|}
  \hline
  \multicolumn{2}{|c|}{Results ( $\lambda_{t}=HIGH$, $c_{Z}=0)$ } \\
  \hline
  $V(t,u)^{*}$ &  $0.0755$ \\
  $V(t,u)^{Cva}$ & $0.2452$\\
  $V(t,u)^{Coll}$ & $0.1116$\\
\hline
\end{tabular}
\end{center}
In the table we have also reported the value function values in the case of no switch allowed, namely $V(t,u)^{Cva}$ is referred to the case $z=1$ (no collateralization) $\forall \: t \in[0,T]$ and $V(t,u)^{Coll}$ is referred to the complement $z=0$ ($\zeta=1$, full collateralization). As one could have expected, the CVA term $V(t,u)^{Cva}$ has the greatest value given the higher default intensities, while the value function $V(t,u)^{*}$ with the contingent switching collateralization allows to minimize the costs (as defined above)  given the greater flexibility of the mechanism with  consistent savings over the funding/collateralization and CVA costs, especially if one considers higher level of contract's notional.\\
In the following ones,  we report similar results as shown above highlighting the variation of the instantaneous switching costs $c_{Z}$ and the effects on value function.
   \begin{figure}[h!]
  \centering
  \includegraphics[  scale=0.55]{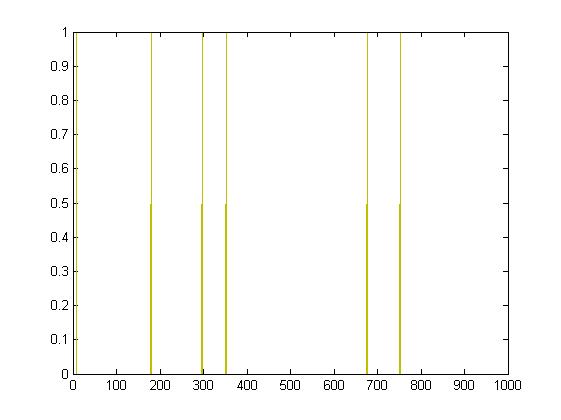}
  \caption{\emph{Graph of the switching indicators $Z=\{0,1\}$ over time $t\in[0,252]$, with $c_{z} = c_{\zeta} = 0.01$} }\label{b}
  \end{figure}
\begin{center}
\begin{tabular}{|c|c|}
  \hline
  \multicolumn{2}{|c|}{Results ( $\lambda_{t}=HIGH$, $c_{z}= c_{\zeta} =0.01)$ } \\
  \hline
  $V(t,u)^{*}$ &  $0.1014$ \\
  $V(t,u)^{Cva}$ & $0.2452$\\
  $V(t,u)^{Coll}$ & $0.1116$\\
\hline
\end{tabular}
\end{center}
\begin{figure}[h!]
  \centering
  \includegraphics[  scale=0.55]{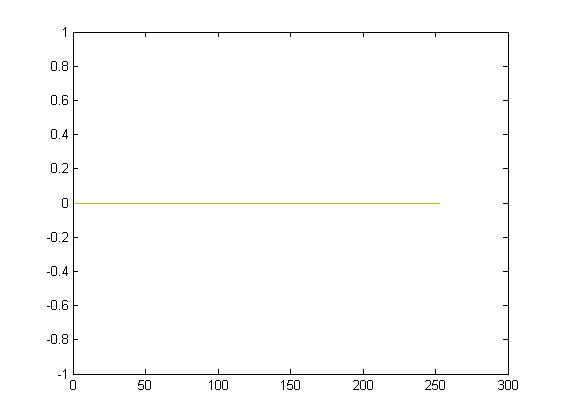}
  \caption{\emph{Graph of the switching indicators $Z=\{0,1\}$ over time $t\in[0,252]$, with $c_{z} = c_{\zeta} = 0.05$} }\label{b}
  \end{figure}
\begin{center}
\begin{tabular}{|c|c|}
  \hline
  \multicolumn{2}{|c|}{Results ( $\lambda_{t}=HIGH$, $c_{z}= c_{\zeta} =0.05)$ } \\
  \hline
  $V(t,u)^{*}$ &  $0.1116$ \\
  $V(t,u)^{Cva}$ & $0.2452$\\
  $V(t,u)^{Coll}$ & $0.1116$\\
\hline
\end{tabular}
 \end{center}

From these graphs and results, we can see that by increasing the instantaneous costs of switching reduces the optimality of switching from a regime to the other as shown in Fig. 5 and 6 in which the vertical lines that indicates the outcome of switching indicators over paths and time are very rare or an horizontal line which indicates zero switches; in particular the tables shows how the value functions tends to increase and converge to the case that we have defined (see in  \emph{Definition 4.5.2} of Mottola (2013) as "\emph{banal switching strategy}", that corresponds also to the value function of the case of not switching $V(t,u)^{Coll} = V(t,u)^{*}$.\\
We made also some trial by increasing the value of $\delta$; in general this increases the cost functions and reduce the convenience to switch but with different setting of the switching costs the effect on the value function is not obvious. So, it is important to study each case carefully.
From a risk management point of view, it is relevant to know or to estimate in advance  the instantaneous switching costs ($c_{z},c_{\zeta} $) in order to determine the optimal savings through the flexibility of switching mechanism.\\\\\\


\item[\textbf{Results in "LOW CASE"}.] In this case the parametrization set is the following:\\
       - $\lambda_{t} = $ "LOW";\\
       - $c_{z} =0 $ and $c_{\zeta} = 0$;\\
       - $\delta = 0$;\\
       - $Z=\zeta =0 $ in $T$, at maturity;\\
       - $N_{paths} = 1000$.\\
   Keeping the other parameters as set above, we get the following main results.\\

 \begin{figure}[h!]
  \centering
  \includegraphics[  scale=0.55]{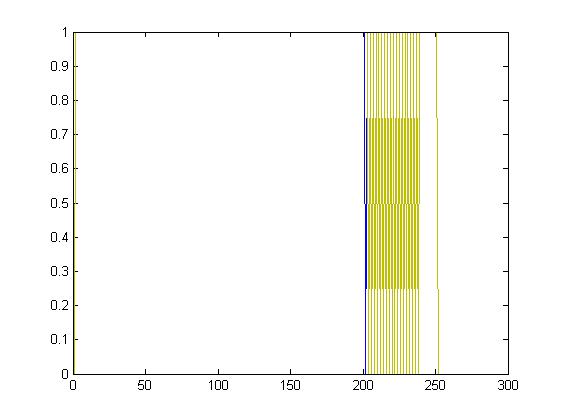}
  \caption{\emph{Graph of the switching indicators $Z=\{0,1\}$ over time $t\in[0,252]$, with $c_{z} = c_{\zeta} = 0.0$} }\label{b}
\end{figure}
   In this case, although the switching costs $c_{Z}$ are null, we can see from Fig. 7 that the lower default intensities $\lambda_{t}$ determine the optimality for counterparty to switch only in the last part of the time domain where the expected cost functions of the two regimes turn to be really near path by path. This can be also easily checked by the plot of the minimum number of remaining switches through time (registered over all the paths) (Fig. 8).

 \begin{figure}[h!]
  \centering
  \includegraphics[  scale=0.8]{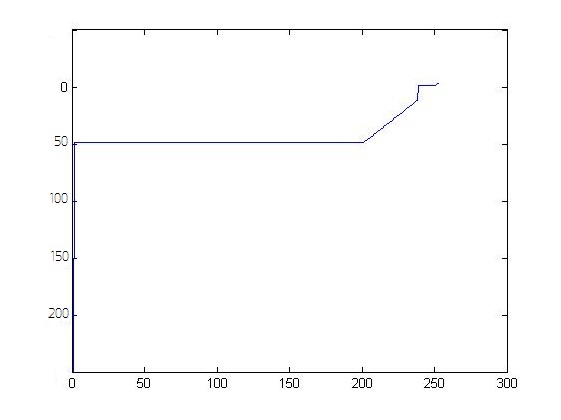}
  \caption{\emph{Graph of the minimum number of remaining switches for all the paths $N_{paths}$ over time $t\in[0,252]$} } \label{b}
\end{figure}

  If one tries to set significantly positive instantaneous switching costs $c_{Z}>0$, no switches take place anymore getting similar results as shown in Fig. 6.\\
  This behavior can be  also verified from the results of the value function in this case.

\begin{center}
\begin{tabular}{|c|c|}
  \hline
  \multicolumn{2}{|c|}{Results ( $\lambda_{t}=LOW$, $c_{z}= c_{\zeta} =0)$ } \\
  \hline
  $V(t,u)^{*}$ &  $0.0117$ \\
  $V(t,u)^{Cva}$ & $0.0153$\\
  $V(t,u)^{Coll}$ & $0.1116$\\
\hline
\end{tabular}
\end{center}
From the table, we note that the model value function $V(t,u)^{*}$ and that without possibility to switch from the $CVA$ regime $V(t,u)^{Cva}$ are almost equal, while the running costs  of the collateralized regime are higher and equal to the case $c_{Z}=0$ of the HIGH case results. Here in fact what changes are just the default intensities which have much lower drift and volatility than the latter case. Similar reasoning and careful check has to be done by changing the other parameters of the model.
\end{description}
From the results just showed, we can say that:
\begin{itemize}
  \item the value function results show a clear dependence on counterparty default intensities which impact on CVA cost function, given the other parameters; of course varying also the collateral/funding costs we would get different values of $V(t,u)^{*}$, as we expected from the theory;
  \item we have also shown that the number of switching times depends on how far are over the paths the cost functions and the related \emph{conditional expectation from continuation}. If these terms are near we verify period of \emph{switching activity} especially if the instantaneous switching costs are null; instead by increasing these costs (both) we get that the value function tends towards a \emph{banal solution}\footnote{We recall this is defined in definition 4.5.2 of Mottola (2013).};
  \item we also mention that the higher number of switching increase in general the value of the contingent collateralization of switching type as one would expect in the sense that a similar type of optionality should be valued more as long as the switches are expected to be "convenient" over time. This also implies that this type of contingency should be valued more and can be an optimal way to manage counterparty risk in cases in which counterparties default intensities are high an in ''volatile'' market conditions.
\end{itemize}


\section{\large Conclusions and further research.}

The present research work has been focussed on the study and the analysis of general \emph{defaultable OTC contract} in presence of the so called \emph{contingent CSA}, that is a counterparty risk mitigation mechanism of switching type. The underlying economic idea and objective has been to show that standard/regulatory  collateralization mechanism are useful to reduce the default losses but they imposes to take in account the cost of collateral and funding, in contrast, by the other side, to the (bilateral) CVA costs. In this sense we have been lead to analyze this problem from the \emph{risk management and optimal design } point of view, taking the perspective of just one party of the contract. By tackling this task, we have formulated a \emph{stochastic switching control model} in which the counterparty can optimally decide to switch from zero to perfect collateralization over the life of the underlying contract. We have analyzed the main possible approaches - analytical/PDE and stochastic/BSDE - in order to show the existence of the solution and of the optimal switching strategy, finding in the stochastic one the most suitable to tackle our highly non linear and recursive model.\\
We have also studied the numerical approaches for the solution coming to the definition of an algorithm based on an \emph{iterative optimal stopping} approach combined with the \emph{Longstaff-Schwartz} simulation method.\\
The results  in the specific case of a defaultable interest rate swap, show clearly - especially when counterparty default intensities are high and volatile - the relevance of the switching mechanism in reducing the overall costs respect to the \emph{perfect} and \emph{zero collateral} cases, that is in absence of contingent CSA. Of course further researches are  necessary in order to remove some of the working assumptions 
and to generalize the problem objective functional or also the numerical solution approach.\\
A tough but interesting case is when the contingent mechanism becomes bilateral and the strategic interaction is allowed between the counterparties and the \emph{price-hedge problem} for this type of contracts with contingent collateralisation, which will be definitely objects of future works.


\newpage
\section*{\large References}

\noindent  T.R. Bielecki, S.Crepey, S. Jeanblanc \& B. Zagari (2011) Valuation and Hedging of CDS Counterparty Exposure in a Markov Copula Model, \emph{International Journal of Theoretical and Applied Finance}, Forthcoming 1-38.\\

\noindent T.R. Bielecki, i. Cialenko \& I. Iyigunler (2011) Counterparty Risk and the Impact of Collateralization in CDS Contracts. available at \texttt{arxiv.org}.\\






\noindent  D. Brigo, A. Capponi, A. Pallavicini \& V.Papatheodorou (2011), Collateral Margining in Arbitrage-Free Counterparty Valuation Adjustment including Re-Hypotecation and Netting, available at \texttt{arxiv.org}, pages 1-39.\\

\noindent  D. Brigo \&  F. Mercurio (2006) \emph{Interest Rate Models - Theory and Practice}, Springer Verlag.\\

\noindent D. Brigo \& A. Pallavicini (2006)  Counterparty risk and Contingent CDS valuation under correlation between interest-rates and default, available at \texttt{ssrn.com}, pages 1-19.\\

\noindent  D. Brigo, A. Pallavicini \& A. Papatheodorou (2010) Bilateral counterparty risk valuation for interest-rate products: impact of volatilities and correlations, available at \texttt{arxiv.org}.\\

\noindent D. Brigo, A. Pallavicini \& D. Parini (2011)  Funding Valuation Adjustment: a consistent framework including CVA, DVA, collateral,netting rules and re-hypothecation,  available at \texttt{arxiv.org}.\\

\noindent M. Carmona \& M. Ludkovski (2010) Valuation of Energy Storage: An Optimal Switching Approach,  \emph{Quantitative finance}, 10(4), 359-374.\\




\noindent S. Crepey (2012) A BSDE Approach to Counterparty Risk under Funding Constraints. Forthcoming.\\


\noindent B. Djehiche \& S. Hamadene (2007) On a finite horizon Starting and Stopping Problem with Default risk,  Preprint.\\

\noindent  B. Djehiche,  S. Hamadene \& A. Popier (2008) A Finite Horizon Optimal Multiple Switching Problem, \emph{Siam Journal Control Optim.}, 48(4), 2751-2770.\\


\noindent  D. Duffie \& M. Huang (1996) Swap Rates and Credit Quality, \emph{The Journal of Finance}, 51(3), 921-950.\\

\noindent N. El Karoui, C. Kapoudjian, E. Pardoux, S. Peng \& M. C. Quenez (1997)  Reflected solutions of backward SDE's and related obstacle problems for PDE's. \emph{The Annals of probability} 25(2), 702-737.\\

\noindent H. Fleming \& H. M. Soner (2008) \emph{Controlled Markov processes and viscosity solutions}. Springer Verlag.\\

\noindent  M. Fujii, Y. Shimada \& A. Takahashi (2010) Collateral Posting and Choice of Collateral Currency,  \emph{Implications for Derivative Pricing and Risk Management}, Technical report, available at \texttt{arxiv.org}.\\

\noindent M. Fujii  \& A. Takahashi (2011) Derivative Pricing under Asymmetric and Imperfect Collateralization and CVA, \emph{Quantitative Finance}, 13(5), 749-768. \\

\noindent  P. Glassermann (2004) \emph{Monte Carlo methods in financial engineering}. Springer Verlag. \\

\noindent J. Gregory (2010) \emph{Counterparty credit risk: The New Challenge for Global Financial Markets.} Wiley Financial Series.\\

\noindent  S. Hamadene \& J. Zhang (2010) Switching problem and related system of reflected backward SDEs,  \emph{Stochastic processes and their applications}, 120(4), 403-426.\\

\noindent M. Jeanblanc \&  Y. Le Cam (2008). Reduced form modelling for credit risk, available at \emph{Default-Risk.com}.\\

\noindent M. Johannes \& S. Sundaresan (2003) Pricing collateralized swaps, Technical report, available at \texttt{arxiv.org}.\\

\noindent J. Karatzas \& S. Shreve (1998)  \emph{Brownian motion and stochastic calculus}. Springer-Verlag.\\

\noindent  F. A. Longstaff \& E. S. Schwartz (2001) Valuing american option by simulations: a simple least squares approach, \emph{Review of financial studies}, 14(1) 113-147.\\

\noindent  G. Mottola (2013), \emph{Switching type  valuation and design problems in general OTC contracts with CVA, collateral and funding issue}. PhD Thesis, School of Economics, Sapienza University of Rome\\


\noindent  B. Oksendal \& A. Sulem (2006) \emph{Applied stochastic control of jump diffusion}. Springer-Verlag.\\

\noindent H. Pham (2009) \emph{Continuous-time Stochastic Control and Optimization with Financial Applications}. Springer Verlag.\\

\noindent  H. Pham \&  V. L.  Vath (2007) Explicit solution to an optimal switching problem in the two-regime case,  \emph{SIAM Journal on Control and Optimization}, 46(2), 395-426.\\

\noindent V. Pieterbarg (2011) Funding beyond discounting: collateral agreements and derivatives pricing, \emph{Risk}, pages 1-15.\\


\noindent  J. Yong \& X. Y. Zhu (1999) \emph{Stochastic controls. Hamiltonian systems and HJB equations.} Springer-Verlag. \\


\end{document}